\RequirePackage{fix-cm}
\documentclass[twocolumn,epjc3]{svjour3}  
\smartqed  
\RequirePackage{graphicx}
\RequirePackage{booktabs}
\usepackage{multirow}
\usepackage{bm}
\usepackage{float}
\usepackage{algpseudocode,algorithm}
\usepackage{xcolor}
\usepackage{cryptocode}
\usepackage{enumitem}
\usepackage{url}
\usepackage{placeins}

\RequirePackage{caption, float}

\newcommand{\chGame}{\mathcal{C}}
\newcommand{\secret}{s_t}
\newcommand{\graycomment}[1]{\textcolor{gray}{\scriptsize\texttt{#1}}}

\journalname{International Journal of Information Security, }
\begin{document}

\title{Synthetic Data: Revisiting the Privacy-Utility Trade-off
}


\author{Fatima Jahan Sarmin\thanksref{addr1}
        \and
        Atiquer Rahman Sarkar\thanksref{addr1, e1} 
        Yang Wang\thanksref{addr2}
        \and
        Noman Mohammed\thanksref{addr1}
}

\thankstext{e1}{e-mail: sarkarar@myumanitoba.ca}

\institute{Dept. of Computer Science, University of Manitoba, Canada \label{addr1}
           \and
           Dept. of Computer Science and Software Engineering, Concordia University, Canada \label{addr2}
}

\date{Received: xx-xx-2024 / Accepted: xx-xx-2024}

\maketitle

\begin{abstract}
Synthetic data has been considered a better privacy-preserving alternative to traditionally sanitized data across various applications. However, a recent article challenges this notion, stating that synthetic data does not provide a better trade-off between privacy and utility than traditional anonymization techniques, and that it leads to unpredictable utility loss and highly unpredictable privacy gain. The article also claims to have identified a breach in the differential privacy guarantees provided by PATE-GAN and PrivBayes. When a study claims to refute or invalidate prior findings, it is crucial to verify and validate the study. In our work, we analyzed the implementation of the privacy game described in the article and found that it operated in a highly specialized and constrained environment, which limits the applicability of its findings to general cases. Our exploration also revealed that the game did not satisfy a crucial precondition concerning data distributions, which contributed to the perceived violation of the differential privacy guarantees offered by PATE-GAN and PrivBayes. We also conducted a privacy-utility trade-off analysis in a more general and unconstrained environment. Our experimentation demonstrated that synthetic data indeed achieves a more favorable privacy-utility trade-off compared to the provided implementation of \textit{k}-anonymization, thereby reaffirming earlier conclusions.

\end{abstract}

\section{Introduction}
\label{intro}
\par In recent years, the importance of data in various fields has increased exponentially. For various data analytics and machine learning applications, a thorough analysis and a robust model training  require a large amounts of high-quality data which provides a diverse and comprehensive representation of real-world scenarios. In many cases, obtaining such data can be difficult or even impossible due to cost limitations, privacy concerns, or simply the lack of available data. Regulatory laws such as the General Data Protection Regulation (GDPR) in the EU and the the Health Insurance Portability and Accountability Act (HIPAA) in the USA often restrict the sharing and publishing of certain privacy-sensitive data in their original form. Restrictions like these led to the research area of privacy-preserving data publishing (PPDP) \cite{fung2010privacy}. Over the decades, many data-sharing models have been proposed and practised. From the early statistical disclosure control to the more recent development of differential privacy, the goal of various privacy-preserving data publishing models is to provide practical utility while protecting individuals' privacy \cite{nist2023deid}. A well-implemented PPDP technique is supposed to significantly reduce the risk of re-identification (where data is abused to identify individuals who contributed to that dataset). However, it is challenging to achieve complete anonymity while maintaining the usefulness of data. Additionally, as technology evolves, so do methods for de-identification and re-identification, highlighting the need for ongoing research and best practices in data privacy.
\sloppy
\par Recent breakthroughs in machine learning techniques such as the generative adversarial networks (GANs) \cite{goodfellow2020generative}, variational auto-encoders (VAEs) \cite{kingma2013auto}, and large language models \cite{patel-NVIDIA-24} have  reignited the interest in synthetic data generation. Synthetic data generation (SDG) is now considered to be of significant potential in both data augmentation and privacy preservation \cite{jordon2022synthetic,cai2022GANs,hernandez2022synthetic}. 
SDG involves creating an artificial dataset that captures the statistical properties of the original data while ensuring that the actual records remain undisclosed. Synthetic data, by design, does not directly expose sensitive information, making it a compelling solution for privacy preservation. As a potential solution to data sharing limitations, it has garnered significant attention from academia and industry \cite{patel-NVIDIA-24,hernandez2022synthetic,elemam2020practical,mcclure2012differential,liu2022synth}. SDG has been examined and assessed as a potential contender for high-quality privacy-preserving data publishing in multiple studies. 
For example, El Emam et al. \cite{elemam2020eval} introduced and applied a methodology to assess identity disclosure risks in fully synthetic data using a COVID-19 cases database from Canada, concluding that SDG substantially mitigates the risks of revealing meaningful identities. Zhang et al.\cite{zhang2021privsyn} developed a differentially private tabular synthetic data generation technique that captures the correlations and can handle 100 attributes with a large domain size $(>2^{500})$. A recent report from National Institute of Standards and Technology (NIST) titled “De-Identifying Government Datasets: Techniques and Governance”\cite{nist2023deid} has considered synthetic data among the “best practices developed over the past several decades” and advised the agencies to consider using synthetic data (in the conclusion section, \textit{Advice for Practitioners}).

On the other hand, \textit{k}-anonymity \cite{samarati1998protecting} provides privacy by ensuring that each record is indistinguishable from at least \textit{k}-1 other records with respect to certain identifying attributes known as quasi-identifiers. However, despite its simplicity, \textit{k}-anonymity has several critical limitations that undermine its effectiveness in real-world applications. Achieving \textit{k}-anonymity often requires significant data generalization and suppression, which can lead to a substantial loss of data utility, rendering the anonymized data less useful for analysis \cite{aggarwal2005k,elemam2008k_anon}. Additionally, it has been demonstrated that \textit{k}-anonymity can skew dataset results or introduce bias in outcomes \cite{Angiuli2015kSkew,slijepvcevic2021k}. Moreover, \textit{k}-anonymity is vulnerable to homogeneity attacks, where all records in an equivalence class share the same sensitive value, and background knowledge attacks, where an attacker uses additional information to re-identify individuals \cite{machanavajjhala2007diversity}.

\par However, a recent paper presented at USENIX-2022 \cite{stadler2022synthetic} (henceforth  referred to as ‘SDR’, named after the accompanying code repository \cite{ground2021git}) claimed that “\textit{synthetic data either does not prevent inference attacks or does not retain data utility}”, “\textit{synthetic data does not provide a better trade-off between privacy and utility than traditional anonymization techniques}”, and synthetic data causes \textit{“unpredictable”} utility loss and  \textit{“highly unpredictable”} privacy gain. In their article, the authors introduced a new membership-privacy game and conducted a comparative analysis focusing on the privacy-utility trade-off between a custom \textit{k}-anonymization approach versus synthetic dataset generation leading to those contradicting conclusions. They also claimed to have found a violation of the differential privacy guarantee in the implementation of PATE-GAN\cite{jordon2018pate} and PrivBayes\cite{zhang2017privbayes}. If these findings are valid, it means that much of the effort invested in leveraging synthetic data for privacy protection is in vain, as it may be just as effective to release the original data after applying simple \textit{k}-anonymity and outlier capping. 

\par\textbf{Contributions.} The trade-off between data utility and privacy remains a key consideration when adopting the synthetic data publishing approach. While synthetic data provides a layer of data protection, it is essential to evaluate its utility in specific analysis or machine learning tasks to ensure that meaningful insights are still attainable. In this evolving data privacy and utility landscape, striking the right balance has become imperative for responsible and effective data management. As it is a relatively new domain, researchers are continuously introducing newer privacy and utility metrics. It is important that whenever new metrics or methods are introduced, the research community investigates and validates their correctness and effectiveness. This is especially important if the newly introduced metrics and methods claim to refute or invalidate prior findings and claims. In this regard, our contributions are as follows. 

\par\textbf{Firstly}, we thoroughly investigated the novel privacy metric and the utility measurement approach introduced in SDR \cite{stadler2022synthetic}. We identified several important privacy and utility measurement characteristics of their experimentation that limited the scope and applicability of their results. In particular, the experimental setup of the privacy game in SDR is very unique. When creating member and non-member seed datasets for the membership privacy game in their experiment, the member and non-member datasets came from two distinctly different distribution. Moreover, while testing the performance of the membership attacker, the test datasets representing the ‘non-member’ cases were devoid of any non-member outliers. Consequently, generalizing any privacy and utility conclusions from this experimental setting should be done with proper reservation. We demonstrated that when the existence of representative non-member outliers are considered, the attacker's success of the specific outlier-focused membership-inference game drops significantly, giving the target much room for deniability.

\textbf{Secondly,} SDR's new privacy game claimed to find a violation of the “differential privacy guarantee” provided by PATE-GAN\cite{jordon2018pate} and PrivBayes\cite{zhang2017privbayes}. SDR was uncertain about the cause of this violation and speculated that it may stem from the implementation aspects of PATE-GAN and PrivBayes. We found the cause. The execution of SDR's privacy game does not satisfy a crucial precondition (Yeom et al.\cite{yeom2018privacy}), resulting in the supposed violation of the privacy guarantee promised by PATE-GAN and PrivBayes. We have provided the evidence of this violation in Section \ref{secFairComp}.

\par \textbf{Finally}, we performed a new privacy and utility evaluation of synthetic data under a general environment so that the findings are generally applicable. We ensured that the competing anonymization techniques are operating under the same environment. Using the same dataset (the Texas healthcare dataset), we evaluated the trade-off between utility and privacy using existing statistical metrics, machine learning applications, and MIA attacks. Our results show the privacy-utility trade-off across different levels of privacy and utility requirements in the context of synthetic data creation and SDR's custom \textit{k}-anonymization approach. In this evaluation, utility aspect showed a generally predictable trend. For the privacy aspect, we observed that privacy-focused synthetic data generators, such as PATE-GAN and PrivBayes, exhibit a predictable trend in privacy protection. However, with \textit{k}-anonymity and other non-privacy-focused synthetic data generators, the attacker's advantage sometimes does not follow a clear pattern. This phenomenon was more pronounced in the case of \textit{k}-anonymity compared to other synthetic data generation models. 

\par Besides, although \textit{trade-off} is mentioned in previous studies, any figure for visualizing the privacy-utility trade-off for different levels of protection offered by various configurations of anonymization was missing. For example, in SDR, the utility and the privacy of \textit{k}-anonymization were measured for only one value of \textit{k} (i.e., \textit{k}=10), leaving the comparison incomplete (e.g., is there a \textit{k} where a competing method performs better/worse?). We have included a visual reporting template which depicts the quantitative interpretations of the privacy-utility trade-off simultaneously, which can aid the data publishers in selecting the proper sanitization technique at the desired privacy level  (Section \ref{reportingTemplate}). 

\par The rest of the article is organized as follows. Section \ref{secBackground} provides the background information necessary to contextualize the privacy-utility trade-off in PPDP. Section \ref{secFairComp} explains the scope limiting factors of SDR's game, providing both theoretical explanation and empirical evidences. (For a brief description of SDR's approach, including the evaluation metrics and their novel privacy game, see Appendix D.) Section \ref{secMethod} outlines our methodology for evaluating the privacy-utility trade-off using statistical utility metrics and existing membership inference attacks. In Section \ref{secResults}, we present the results of our re-investigation and demonstrate the application of a visual reporting template for a simultaneous comparison of different privacy-utility trade-offs offered by various data publishing techniques. Finally, Section \ref{secConclusion} concludes the paper and suggests directions for future research.

\section{Background}\label{secBackground}
When sharing a dataset for analysis, striking the right balance between privacy and utility is crucial, as enhanced privacy protection measures can impact the usability of the data. As a result, data owners must consider three key factors to effectively manage this trade-off. These factors include: (1) selecting privacy models and algorithms, (2) quantifying privacy risks, and (3) evaluating data utility. Next, we will provide an overview of these factors. It places greater emphasis on describing \textit{k}-anonymity, differential privacy (DP), and membership inference attack (MIA), as these concepts are essential for understanding the subsequent sections.

\subsection{Privacy Models and Algorithms}\label{subsecModel}
Preserving data privacy before sharing involves minimizing the potential for unintended information disclosure. This objective can be attained by employing a range of privacy models and algorithms, which can be broadly classified into two categories: (1) traditional anonymization techniques and (2) synthetic data generation.

\subsubsection{Traditional anonymization} \label{subsubsecTradAnon}
Traditional anonymization models modify the records of the original dataset and establish a link between the anonymized and the original dataset. \textit{k}-anonymity \cite{samarati1998protecting,sweeney2002k}, \textit{l}-diversity \cite{machanavajjhala2007diversity}, and \textit{t}-closeness \cite{li2006t} are popular models of traditional anonymization. In its simplest form, \textit{k}-anonymity can be defined as follows.

Let \textit{D} be a dataset consisting of \textit{n} tuples, where each tuple \( t \in D \) represents a data record of an individual. Suppose \textit{D} consists of \textit{m} attributes: \( A_1, \dots, A_m \). A quasi-identifier \( Q_D \) of \textit{D} is a set of attributes \( \{A_i, \dots, A_j\} \subseteq \{A_1, \dots, A_m\} \) that can potentially identify individuals (e.g., \(\{age, gender, zip code\}\)). For a given dataset \( D \), let \( G \) be a group of tuples that share the same values for the quasi-identifiers \( \{A_i, \dots, A_j\} \). The dataset \( D \) is \( k \)-anonymous if:
$$|G| \geq k, \quad \forall G \subseteq D$$
where \( |G| \) denotes the size of the group \( G \).
\textit{k}-anonymity ensures that individual data records within a dataset cannot be uniquely distinguished from at least \( k \) other records, thus protecting the identities of data subjects. \textit{l}-diversity takes this a step further by requiring that each group of \textit{k}-anonymous records also contains at least ‘\textit{l}’ different sensitive attribute values, diversifying the data and making it harder for potential adversaries to infer sensitive information. \textit{t}-closeness addresses attribute disclosure by maintaining the distribution of sensitive attribute values within each group to be consistent with the overall dataset distribution. 

Given a privacy model, various anonymization techniques are employed to transform the original dataset into a version that satisfies the privacy model. Generalization and suppression are two common techniques that are often used for anonymization.  In generalization-based algorithms, attribute values are substituted with more generalized versions, while suppression-based algorithms achieve anonymity by removing the attribute value. Some well-known generalization-based \textit{k}-anonymization algorithms include Mondrian \cite{lefevre2006mondrian}, Incognito \cite{lefevre2005incognito}, and Datafly \cite{sweeney1998datafly}. Some examples of suppression-based anonymization techniques include record suppression \cite{samarati2001protecting,bayardo2005data}, value suppression \cite{wang2007handicapping}, and cell suppression \cite{meyerson2004complexity}. In these techniques, algorithms can suppress entire records, all instances of a specific value within a table, or select instances of a particular value in a table, respectively.

\subsubsection{Synthetic data generation}\label{subsubsecSDG}
Synthetic data is typically generated to replicate the characteristics and patterns of real data. Unlike traditional anonymization, there is no one-to-one relationship between real and synthetic data. 
Synthetic data generation is not a new concept \cite{rubin1993statistical}; however, it has gained renewed interest in the post deep learning era. Various methods are available for creating synthetic data, which can be  broadly classified into two groups: statistical methods and deep learning-based methods. Statistical techniques aim to replicate the statistical properties and relationships found in actual datasets (e.g., Bayesian networks \cite{young2009using}, Hidden Markov models \cite{ngoko2014synthetic}). On the other hand,  deep learning models acquire the ability to determine the relevant attributes through a stochastic training process (e.g., GANs \cite{goodfellow2020generative}, VAEs \cite{kingma2013auto}).


Synthetic data can be generated with or without formal privacy guarantees. Differential privacy, an alternative to traditional anonymization, can be integrated into a synthetic data generator to ensure data protection. Differential privacy aims to ensure that the outcome of any analysis does not overly depend on a single data record, providing assurance to every record owner that their participation in a database will not lead to a privacy breach. This approach to privacy offers a precise definition of privacy preservation and standardized evaluation methods. An algorithm \( A \) is \( \epsilon \)-differentially private if, for two datasets \( D \) and \( D' \) differing by one record, the probabilities of all of their corresponding outputs in output space \( S \) are bounded by \( \epsilon \):
$$\Pr[A(D) \in S] \leq e^\epsilon \times \Pr[A(D') \in S]$$

A differentially private algorithm ensures this through careful addition of noise. Two important properties that make \( \epsilon \)-differential privacy an attractive approach are (i) robustness to post-processing and (ii) the sequential composition property. The robustness to post-processing ensures that if an algorithm \( A \) is \( \epsilon \)-differentially private, then any other algorithm that operates on the output of \( A \) is guaranteed to be at least \( \epsilon \)-differentially private. The sequential composition property states that if we make \( t \) queries to an \( \epsilon \)-differential privacy mechanism, with each query being randomized independently, the overall result will be \( \epsilon t \)-differentially private. Conversely, once a model is trained or fine-tuned with \( \epsilon \)-differential privacy, it will remain \( \epsilon \)-differentially private regardless of how many responses to queries are taken from the model. Examples of differentially private synthetic data generators include DPGAN \cite{xie2018differentially}, PATE-GAN \cite{jordon2018pate} and PrivBayes \cite{zhang2017privbayes}. 

\subsection{Privacy Metrics}\label{subPriv}
Privacy metrics provide quantitative means of assessing the level of privacy in a data publishing algorithm. They enable the comparison of different privacy-preserving techniques. Numerous privacy metrics have been proposed in the literature but they are often hard to interpret \cite{wagner2018pvmetrics}. For example, what constitutes an appropriate value for \textit{k} in \textit{k}-anonymization or \textit{$\epsilon$} in differential privacy is not intuitive. An alternative approach is to measure privacy in terms of privacy-attacker's success \cite{dwork2017attacks,yeom2018privacy,giomi2022unified,stadler2022synthetic}. Three of the major attacks are discussed below.

\textbf{Re-identification attacks.}  In the re-identification attack, an adversary attempts to identify individuals within an anonymized/synthetic dataset. This involves correlating quasi-identifiers or other attributes in the anonymized data with external information sources or public datasets to uncover the true identities of individuals \cite{el2011systematic}. The goal is to reveal the actual identity of specific individuals. Another attack named linkage attack is closely related to the re-identification attack. In a linkage attack, the adversary seeks to establish connections or links between records or activities of the same individual across different datasets or contexts \cite{merener2012linkage}. The attacker is interested in showing that two or more records, which may appear unrelated, correspond to the same individual. These attacks can be particularly harmful in scenarios where individuals expect their activities to remain separate and unlinked.

\textbf{Membership inference attack.} A membership inference attack is a privacy attack where an attacker aims to determine whether a specific individual's data was included in the training set of a machine learning model or in the raw dataset of the published anonymized/synthesized data. Since its introduction in \cite{shokri2017membership}, different variations and techniques of the attack has been proposed \cite{stadler2022synthetic,yeom2018privacy,salem2018ml,carlini2022membership}. Here, we present the membership inference attack as a game, inspired by the work of Carlini et al. \cite{carlini2022membership} and Yeom et al. \cite{yeom2018privacy}, which aims to determine whether a specific individual's data was included in the training set of a machine learning model. 

Let $\mathcal{D}_R$ be an underlying real data distribution.

\begin{enumerate}
    \item The challenger samples two disjoint raw datasets: $
    R_{\text{train}}, R_{\text{non-train}} \sim \mathcal{D}_R \quad (R_{\text{train}} \cap R_{\text{non-train}} = \emptyset). $ 
    
    \item The challenger trains a model ML using $R_{\text{train}}$.
    
    \item The challenger chooses a bit \textit{b} uniformly at random from \{0, 1\} and samples a record $\mathcal{X}$. If $\textit{b=0}$, the challenger samples $\mathcal{X}$ from $R_{\text{non-train}}$ or, if $\textit{b=1}$, samples $\mathcal{X}$ from $R_{\text{train}}$. The challenger gives $\mathcal{X}$ to the adversary.
    
    \item The adversary, with access to $\mathcal{D}_R$ and query access to the model ML, predicts \^{b}. If \textit{\^{b}=b}, the adversary wins. Otherwise, the adversary loses.
\end{enumerate}

The attack described above directly targets the training data of a machine learning model. In Appendix D.3, we briefly discuss the attack proposed by SDR \cite{stadler2022synthetic}, which examines published data from a synthetic data generator (or a \textit{k}-anonymizer) to determine whether a particular outlier's data was part of the input dataset for the synthetic data generator (or the \textit{k}-anonymizer). Section \ref{subsecOurMia} presents our adaptation of the MIA attack, inspired by previous works \cite{stadler2022synthetic,yeom2018privacy,carlini2022membership,tf2020tfmia}, focusing on models trained on synthetic or sanitized data. Our adaptation aims to determine if a specific individual's data was included in the seed dataset used by the synthetic data generator or sanitizer. More recently, Ganev \& Cristofaro \cite{ganev2023inadequacy} demonstrated that certain privacy scores provided with synthetically generated datasets can be exploited to infer membership. Although all of these attacks infer membership, their threat models differ. More details will be discussed in the corresponding sections.

\textbf{Attribute inference attack.} An attribute inference attack is an attempt to predict or infer specific sensitive attributes or characteristics of individuals from a dataset, when those attributes are not explicitly present in the published dataset \cite{powar23petsattrib,jia2018attriguard}. However, there exists a fine distinction between what constitutes an attribute inference privacy attack and what is a simple statistical imputation. It has been shown in \cite{jayaraman2022attribute} that some of the previously thought attribute inference attacks are no more than statistical imputation.

\subsection{Utility Metrics}\label{subUTIL}
Before releasing data, it is crucial to evaluate the utility of the datasets. Failing to check the utility aspect after ensuring privacy can be very problematic and even life-threatening (e.g., warfarin dosing\cite{fredrikson2014warfarin}). Various utility metrics have been developed to assess the utility of the processed data. In a broader categorization, these metrics can be divided into two ways: special purpose metrics and general purpose metrics.

\textbf{Special purpose metrics:}
A special-purpose utility metric measures the utility or usefulness of released data for a particular, often narrowly-defined, purpose or application. These metrics are designed to assess how well the data serves a particular use case or meets the requirements of a specific user or system. For example, in healthcare data publishing, a special-purpose utility metric might focus on how well the data supports a particular predictive modeling task. If the data is intended to be used in a machine learning model, usually “train on synthetic/anonymized, test on real” is a commonly practiced approach to measure the utility of the synthetic/anonymized data \cite{hernandez2022synthetic,lefevre2008workload,fung2007anonymizing}. In this case, usually metrics such as the classification/prediction accuracy, precision, recall, F1 score, the area under the curve (AUC), etc. are used to quantify the utility of synthetic/anonymized data.

\textbf{General purpose metrics:}
When data publishers lack knowledge about the specific use cases for the published data, they employ general-purpose metrics to evaluate the similarities and data distribution between the original and sanitized datasets. These metrics vary from simple statistics such as the range, mean, median, standard deviation, to more involved tests such as the Pearson's correlation test, Chi-squared test, to corresponding variable's distribution tests such as the Kullback–Leibler (KL) divergence, Kolmogorov-Smirnov (KS) test,  total variation distance (TVD) and many more \cite{hernandez2022synthetic}. It is possible that a dataset that shows good utility according to a metric is not so good according to another metric focusing on a different aspect \cite{NHS2019synth}. 

\section{Reassessment of SDR} \label{secFairComp}
In this section, we discuss the factors that limited the scope in the novel privacy game in SDR. We demonstrate how changing different aspects of the privacy experiment influence the privacy state of the target record. In Subsection \ref{subsec_flawinGame}, we showed that the privacy experiment in SDR did not consider the cases for population representative outlier in the ‘non-member’ labeled sets even though the raw dataset contained them.  We discuss the theoretical necessity as well as the experimental evidence of this issue in detail. Subsection \ref{subsub_test_set} discusses how a limitation in the test dataset can inflate the attacker's success. It also experimentally shows how the SDR's current attacker faces difficulty inferring non-membership of the target record when a non-member outlier is present in the test dataset. Subsection \ref{subsubsec_violation} provides the evidence of not meeting a critical prerequisite in the privacy game leading to the supposed violation of the differential privacy guarantee promised by PATE-GAN and PrivBayes. Subsection \ref{subsec_unfair} discusses certain choices in the implementation of synthetic data generation and utility comparison that made the playing field uneven among the competing data publishing approaches. In Section \ref{secMethod}, we present our methodology and experimental environment to ensure an equitable comparison among the competing techniques.

\subsection{Accounting for Non-member Outliers} \label{subsec_flawinGame}
El Emam et al.  \cite{elemam2008k_anon} have demonstrated that the actual level of re-identification risk is not contingent on the size of equivalence classes within the sampled dataset. Instead, it is contingent upon the corresponding equivalence class size within the entire population. To illustrate, consider a scenario in which a target record, subjected to ‘\textit{k}=5’ anonymization, ends up in a class of size 5. Although the re-identification risk appears to be 1/5, if there are 100 records within the entire population that match the quasi-identifiers of those five records, the actual risk decreases to 1/100. 

Although re-identification attack and membership inference attack are different in nature (see Section \ref{subPriv}), a similar principle applies. The plausible deniability in the case of a membership inference attack is directly related to the false-positive calls made by an attacker (Attacker's Advantage = True Positive Rate - False Positive Rate). Note that, like in the re-identification attack's case, the risk or an attacker's advantage could be overestimated if the privacy-providing cloaking records in the representative sample/population are ignored and the attacker's evaluation dataset only consists of trivial cases.

\textbf{Empirical evidence from SDR.} The first limitation in SDR's privacy game experiment is that it did not consider the cases where non-member outliers exist in the population. SDR employed random subsampling of very small sizes (e.g., from 50K to 10K, and finally to 1K). This subsampling resulted in the sampled subsets containing no outliers. The ‘member’ labeled sets are created by forcefully inserting the target outlier to the sampled subsets. On the other hand,  the ‘non-member’ labeled sets contained no outliers. This is comparable to an environment where the entire raw population includes only a single outlier—the target outlier. Consequently, the membership classifier effectively becomes a data distribution differentiator. That is the reason why features as simple as the histograms or summary statistics (mean, median and so on) were able to get such good attack success. In the ‘member’ case, the outlier influences the synthetic data generation, causing larger values in the outlier attribute fields. On the other hand, since the non-member cases did not include representative non-member outliers, these synthetic datasets have comparatively lower values in the corresponding attribute fields.

\par
\begin{figure}[t]
  \centering
  \includegraphics[width=1\linewidth]{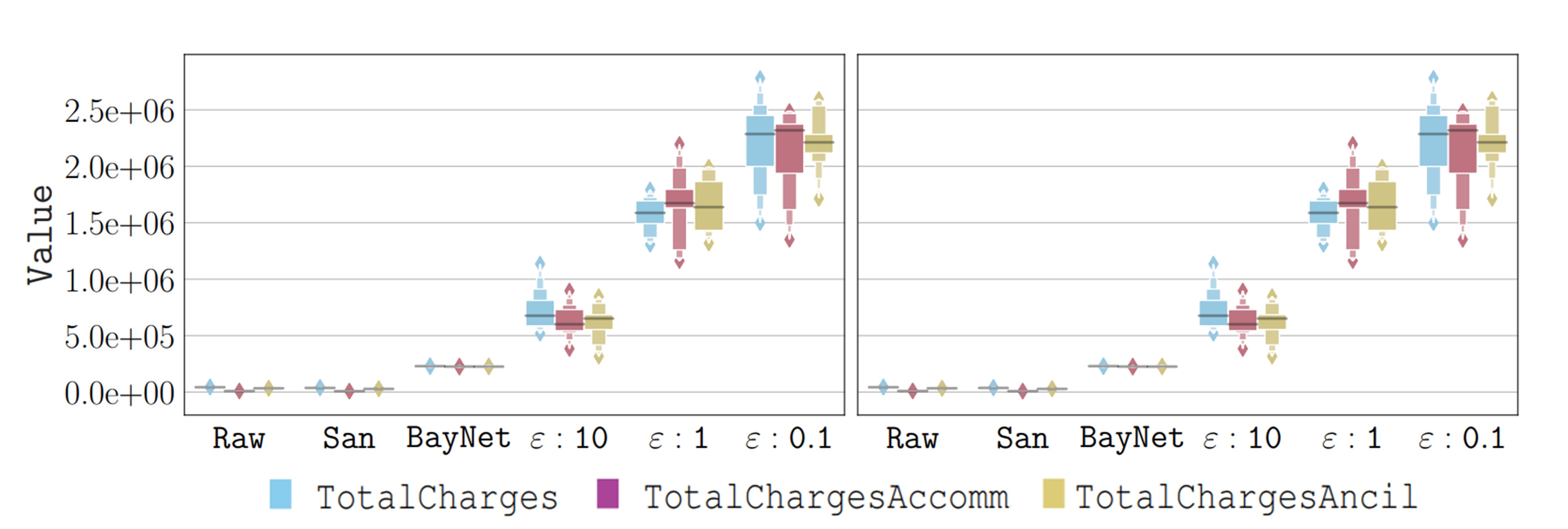}
  \captionsetup{belowskip=0pt} 
  \caption{Mean (left) and median (right) for attributes \texttt{TotalCharges}, \texttt{TotalChargesAccomm}, and
\texttt{TotalChargesAncil} (taken from Fig. 9 of SDR) \cite{stadler2022synthetic}}\label{fig:Fig9GH}
\end{figure}

 We now describe the evidence we found that reveal this limitation. As we investigated the five outliers \cite{ground2021git}, we noticed that they were outliers in the \texttt{TotalCharges}, \texttt{TotalChargesAccomm}, and \texttt{TotalChargesAncil} columns. Unfortunately, the random sampling of the 1K records from the sample of 10K records (that were random-sampled from 50K records) didn't pick any outlier as evident from the mean of those columns of the ‘raw’ datasets in Fig. \ref{fig:Fig9GH} (sourced from Fig. 9 of SDR for the reader's convenience). From the distribution of these 3 columns, we can see that the sample datasets were not representative of the population in terms of outliers. The population means (rounded) for the entire dataset for those columns are 44641, 10049, and 34590, respectively. 

\begin{figure*}[!t]
  \centering
  \includegraphics[width=1\textwidth]{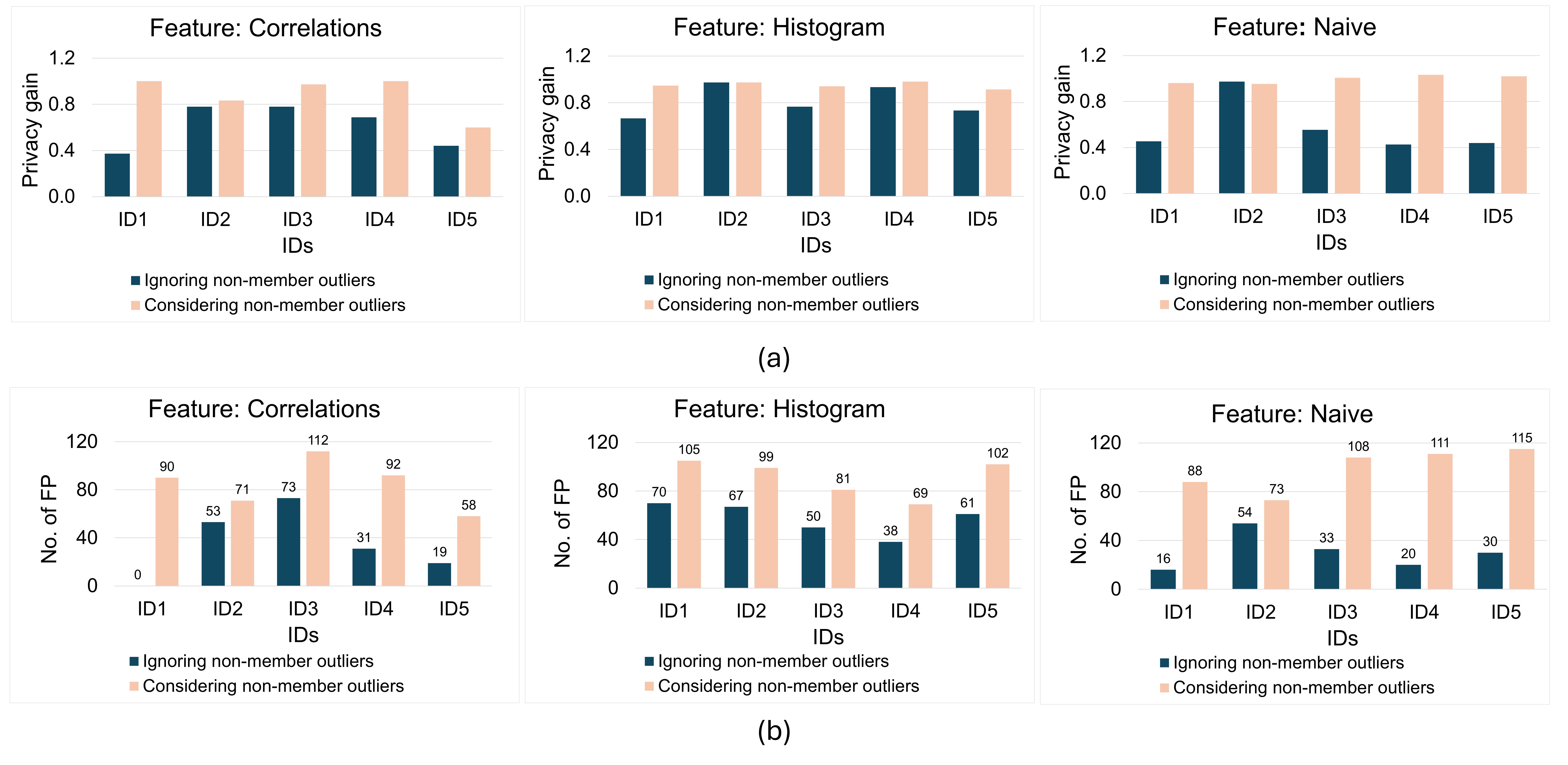}
  \caption{(a). Effect of ignoring and considering non-member outliers on Privacy Gain (BayNet); (b). False positive rate increases when a outlier is present in the non-member set (BayNet). }\label{fig:non_member_PG_FP_increase} 
\end{figure*}

\textbf{Observation.} \textit{Researchers should strive to obtain a representative sample to ensure that the findings are applicable to the real-world data distributions and use cases. If the samples are not representative, the findings may not be generalizable and applicable to the population as a whole.}

\subsection{Limitation of the Test Dataset}\label{subsub_test_set}
The problem described in subsection \ref{subsec_flawinGame} resulted in the attacker being trained with two relatively easily distinguishable classes of datasets: ‘member’ labeled sets associated with a forcefully inserted target outlier and ‘non-member’ labeled sets associated with no outliers. The fact that the attacker's advantage is inflated on the Texas Hospital Discharge dataset would have been detected if the ‘non-member’ test datasets had contained non-member outliers. Unfortunately, the evaluation phase utilized the same subsampling approach as before. Consequently, the same issue associated with the small sampling size resulted in the test datasets devoid of non-member outliers (the outliers that provide the privacy cloak to the target). Fig. \ref{fig:non_member_PG_FP_increase}(a) depicts how a non-target outlier in the seed for ‘non-member’ labeled test datasets fools the trained membership classifier into mis-classifying the datasets as ‘member’, resulting in a significant increase in the privacy gain metric. Fig. \ref{fig:non_member_PG_FP_increase}(b) shows the corresponding increase in the false positive rates.

In Fig. \ref{fig:non_member_PG_FP_increase}, the 150 member samples are generated by 15 synthetic generators trained with datasets sampled from the reference 10K-records dataset (1000+the target outlier) for each generator. The 150 non-member datasets are similarly generated using 15 generators, but a non-member outlier was inserted instead of the target outlier. It should be noted that this modified evaluation phase, in which a non-member outlier was inserted into the test datasets, also fails to address the representativeness issue of the population distribution highlighted in Section 3.1. However, our goal in this section is to demonstrate and expose the inflation of the attacker's success in SDR's approach where non-member outliers are not considered. This modification employs the same sampling and outlier insertion approach from SDR to the non-member cases of the ‘test dataset’ (SDR only applied outlier insertion to the train dataset's ‘member’ cases).

\begin{figure}[btp]
  \centering
  \includegraphics[width=1\linewidth, height=11.5cm]{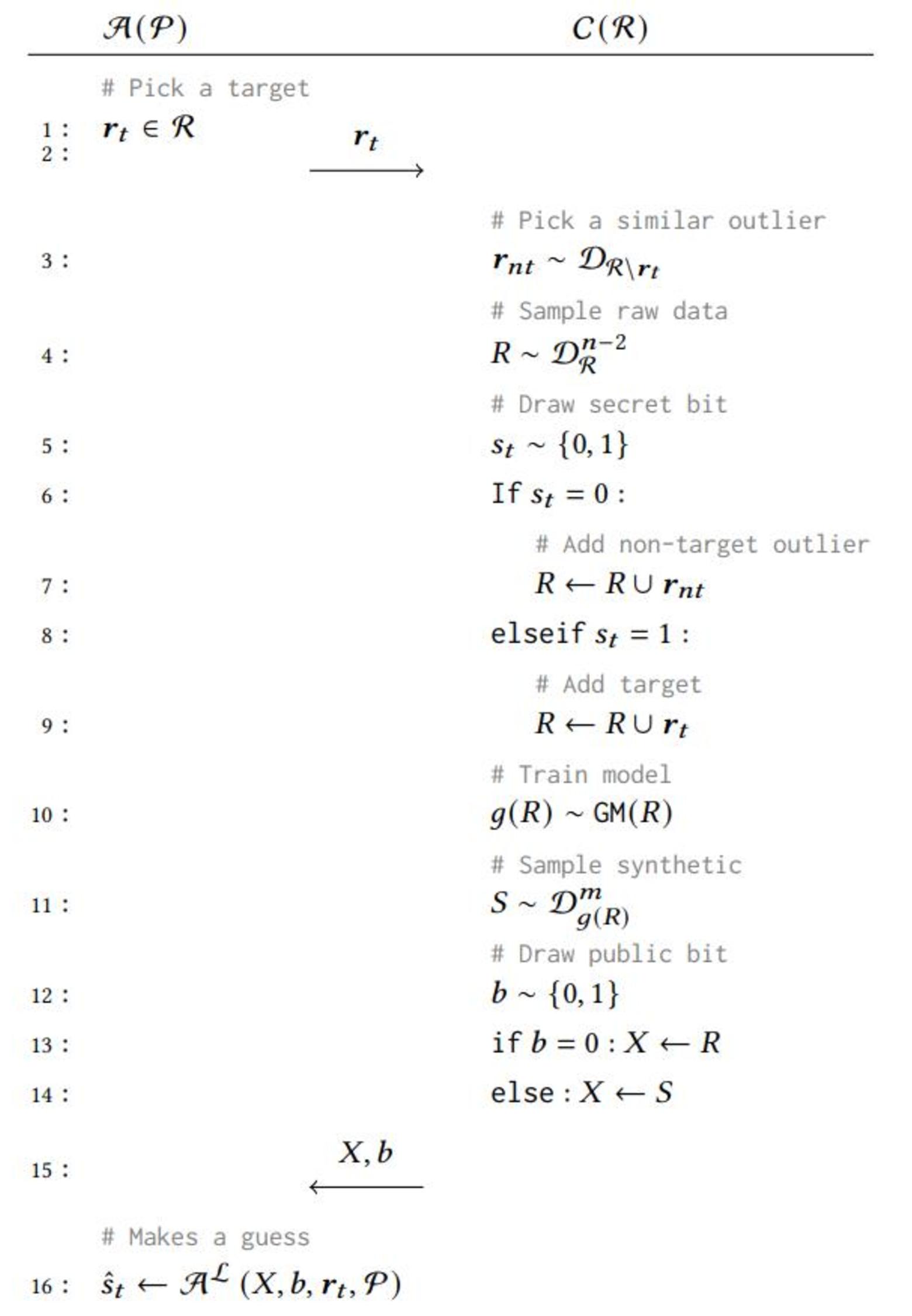}
  \caption{Modified evaluation phase for the membership inference attack.} \label{fig:mia_game}
\end{figure}

The modified evaluation phase for SDR's game is outlined in Fig. \ref{fig:mia_game}. In this modified game, the adversary sends a target record $r_{t}$ to the challenger $\chGame$. The challenger selects a non-target outlier denoted as $r_{nt}$, similar to the target outlier $r_{t}$, from the population $\mathcal{R}$. Then, the challenger samples the raw dataset $R$ from the data distribution $D_\mathcal{R}^{n-2}$. Next, the challenger draws a random secret bit $\secret$. If the generated $\secret$ is 0, the non-target outlier is added to the raw dataset $R$; otherwise, the target outlier is added. Following this step, the challenger trains a generative model $g(R)$ using the training procedure $GM(R)$ to produce synthetic data $S$. Finally, the challenger draws another public random bit $b$. If $b$ is 0, the challenger sends $R$ to the adversary. If $b$ is 1, the challenger sends the generated synthetic data $S$ to the adversary. Upon receiving the dataset, the adversary has to decide, using the received dataset, prior knowledge, the target, and the public bit $b$, whether the secret bit $\secret$ was 0 (for $r_{nt}$) or 1 (for $r_{t}$) in this challenge.

\textbf{Recommendation.} \textit{The test dataset should be carefully chosen to be representative of the target domain because it plays a critical role in assessing the performance and generalization ability. Failure to ensure representativeness in the test dataset may result in biased conclusions, overestimated performance or unwarranted confidence in the effectiveness of the algorithm's or attack's capabilities.}

\subsection{Cause for the Violation of the DP Guarantee}\label{subsubsec_violation}

In subsection 5.1 of SDR, the authors observed that two out of the five outliers achieved a privacy gain (PG) of less than 0.1 in the membership inference attacks. This observation raised concerns as it contradicted the theoretical lower bound offered by differential privacy, as proven by Yeom et al. \cite{yeom2018privacy}. In light of this contradiction, the authors claimed about PATE-GAN and PrivBayes that \textit{“while both models on paper fulfill their formal privacy definitions, their available implementations did not.”}

Upon further analysis, SDR found that the random sampling process was not effectively selecting targets with rare categorical attributes or substantial outliers in numerical attributes. In an effort to address this issue, SDR modified the implementations of PATE-GAN and PrivBayes by explicitly specifying the ranges of numerical attributes and the categories for categorical attributes. Although this adjustment improved the privacy gain considerably, the bound was still violated. The authors stated that \textit{“the remaining gap can likely be explained either by other aspects of the model’s implementation that violate theoretical assumptions and we were not able to find in our analysis”}. 

\textbf{Our findings.} The evidences presented in subsection \ref{subsec_flawinGame} showed us that the member and non-member samples do not represent the same distribution (member samples contained forcefully inserted outliers while non-member samples had no outliers). The question is: are the different distributions responsible for the observed violation of the differential privacy guarantee in SDR? As we look deeper into the concerned bound on the attacker's advantage established by Yeom et al. \cite{yeom2018privacy}, we noticed a violation of an important precondition related to differential privacy. The violated precondition is that the two datasets , which differ by one record, were supposed to have \textit{“\textbf{identical distributions}”} \textit{(Yeom et al. \cite{yeom2018privacy} (page-6): Proof of theorem 1: Bounds from differential privacy)}. We found that, as demonstrated in the previous subsection, even when the two datasets differed by only one record, they represented two completely different distributions, including ranges, means, standard deviations, and peaks. This violation went unnoticed. SDR attempted to resolve this issue by specifying upfront the ranges of numerical attributes and possible values of categorical attributes. However, this solution only partially solved the problem because, although the range was externally fixed, the records inside the dataset did not represent the actual distribution. As a result, it restricts the game's applicability to real-world scenarios.

\textbf{Recommendation.} \textit{Privacy attacks and anonymization techniques often make certain assumptions about their underlying operating environment. Researchers should verify that these attacks or techniques meet the necessary assumptions/preconditions for the reliability of the study.}

\subsection{Variation in Preprocessing Steps}\label{subsec_unfair}
If preprocessing steps vary between competing methods, it becomes challenging to isolate the effects of the anonymization technique itself on the results. Consistency in applying preprocessing steps is crucial to isolate the effects of anonymization techniques on data utility and privacy protection.

SDR claims to implement the sanitization procedure described by the National Health Service (NHS) England \cite{NHS2019synth}. The NHS provided a methodology to create synthetic data and provided an example by creating a synthetic dataset using a Bayesian Network. In that NHS methodology, before using the Bayesian Network, the raw data is pre-processed by replacing geographical data with average demographic data, categorizing highly detailed variables into bands, eliminating specific time details, filtering out infrequent values, capping integer variables, and removing unique values and subsets. In SDR, before performing \textit{k}-anonymization, the preprocessing steps were performed on the raw data, including capping all the values above 95\% quantile to 95\% quantile. However, when generating synthetic data, those preprocessing steps were not performed on the seed raw data, not even when generating synthetic data using a Bayesian Network (despite the NHS report providing a demonstration using a Bayesian Network \cite{NHS2019synth}). 

We believe that the preprocessing of the raw dataset for one approach (i.e.,  \textit{k-}anonymization) while leaving the dataset unprocessed in the competing approach (against the advise by the NHS) does not make a fair comparative study. Capping the outliers in the seed datasets when using \textit{k}-anonymization but keeping them intact in the synthetic data generation process caused an adverse effect in the privacy and utility aspect of the synthetic data. For example, the deviation of the means and medians from the real dataset was used as an utility metric in SDR. Note that the real population was 50K and but the sampled dataset is only (1K + the target outlier). Consequently, the outlier's impact on the mean is even more pronounced in the synthetic dataset. This is unfair because the outlier will clearly affect the mean in the synthetic data generation process, but due to capping it will have little to no impact in the \textit{k}-anonymized dataset. The impact of not capping is more pronounced in the differentially private data generators. As the DP synthetic data generators rely on adding noise to protect the records, having an outlier amidst the records will generate many records between the usual higher end and the outlier value. Adding so many records between the outlier's value and the usual upper-end destroys the distribution of that variable.

\textit{Other concerns:} regarding the mean as a utility metric, as noted in subsection \ref{subsec_flawinGame}, the population mean (rounded) for the \texttt{TotalCharges}, \texttt{TotalChargesAccomm}, and \texttt{TotalChargesAncil} columns in year-2013 were (50K records): 43473, 9572 and 33905 respectively. Therefore, BayNet and PrivBayes were actually closer to the real population mean than the NHS-sanitized data or the unrepresentative raw-samples that did not include any outlier in Fig. \ref{fig:Fig9GH}. Two reasons why the synthetic data produced larger numbers than the unrepresentative samples are (i) due to the modification performed by SDR in PrivBayes and PATE-GAN, and (ii) the forceful inclusion of the outliers.

Furthermore, the differential-privacy literature has already addressed the problem with outliers by either dropping/capping the outlier or, if the outliers are vital and must be kept, ensuring different levels of differential privacy for different groups in the dataset. For example, Lui and Pass \cite{LuiP2015outLpriv} proposed \textit{(k, $\epsilon$)}\textbf{ outlier differential privacy} which requires \textit{($\epsilon$/k)}-DP for \textit{k-}outliers and \textit{$\epsilon$}-DP for the other individuals. They also proposed staircase outlier privacy, which involved more groups with different DP requirements for each group. In Section \ref{secResults}, we have shown the utility and privacy measurement experimentation where outliers are capped in the seed raw datasets to the synthetic data generation process.

SDR also claimed that synthetic datasets \textit{“do not preserve the fine-grained statistical patterns needed for outlier analysis.}” However, there are techniques in the outlier analysis literature that use this particular property of general trend to distinguish the outlier from the population. For example, Mayer et al. \cite{mayer2020privacy} leveraged an autoencoder neural network for detecting outliers and novel data points. This neural network comprises an encoder network responsible for reducing input data dimensions and a decoder network focused on reconstructing the input data. Although dimensionality reduction leads to information loss, the learning objective encourages the preservation of common information among most training samples. This approach enables the identification of outliers containing rare information by assessing the model's loss. To determine whether a given data sample is an outlier using autoencoders, any sample exhibiting a loss exceeding a learned threshold is considered an outlier or novelty. The authors \cite{mayer2020privacy} noted that while anomaly detection is inherently challenging, synthetic data demonstrated comparable effectiveness to models trained on the original data in specific scenarios. Their evaluation was conducted using the “Credit card fraud” dataset available at Kaggle.com. More recently, Hu et al. \cite{hu2024sok} experimented with image data and demonstrated that the pessimistic conclusion by Stadler et al.—that achieving strong privacy necessitates a significant sacrifice in utility—is not entirely applicable in the image domain. Through their experimentation with DP-MERF, they observed that achieving both high utility and strong privacy in synthetic data generation is feasible and simply requires further exploration to uncover such solutions.
 
\section{Methodology and Experimental Setup}\label{secMethod}
This section describes our experimental framework for assessing the privacy-utility trade-off between the \textit{k}-anonymization-based sanitization \cite{stadler2022synthetic} and some synthetic data generation techniques. We experimented with three datasets: (i) the Texas healthcare and (ii) the Adult income datasets, which were chosen because they were used in the SDR article, and the (iii) Credit Card Churning dataset (henceforth referred to as the credit card dataset). The datasets have been described in Appendix B. All three are tabular datasets with a mix of numerical and categorical attributes. For brevity and clarity, as done in the SDR article, we present the results from the Texas dataset in the main body, comparing \textit{k}-anonymization with four synthetic data generators: BayNet \cite{ping2017datasynthesizer}, PrivBayes \cite{zhang2017privbayes}, PATE-GAN \cite{jordon2018pate}, and TabDDPM \cite{kotelnikov2023tabddpm} (henceforth referred to as DDPM). The experimental results using the other two datasets are provided in Appendix C; however, they are discussed in the Results section. Below, we describe our methodology and experimental setup.

We believe that a balanced comparative study should provide an equal setting for all the competing methods. To address the fairness concern mentioned in subsection \ref{subsec_unfair} (specifically, the preprocessing steps recommended by the NHS-England being applied to \textit{k}-anonymization but not to synthetic data generation), we also applied outlier capping at the 95th percentile to the seed dataset for the generation of synthetic data. For \textit{k}-anonymization, SDR used k = 10. We extended the experiments and used four different values of \textit{k} (\textit{k} = 5, 10, 15, 20) to generate different datasets. Measuring the utility and privacy for multiple values of \textit{k} will enable us to understand the utility trade-off for different levels of privacy. From the literature, it is known that as \textit{k} increases, privacy usually increases while utility usually decreases. We used four distinct bin values (bin = 2, 5, 10, 25) in the ‘bin’ parameter of the BayNet algorithm. This parameter controls the granularity of the synthetic data.  For differentially private generation, SDR experimented with $\epsilon$ = 0.1, 1, and 10. We additionally used 15 to observe the trade-off ($\epsilon = $ 0.1, 1, 10, and 15). For DDPM, the authors recommended diffusion timesteps between 100 and 1000 \cite{kotelnikov2023tabddpm}. We experimented with timesteps of 100, 500, 1000, and 1500.

For privacy measurement, we adapted a membership inference attack \cite{tf2020tfmia} and also utilized SDR's privacy game. We used the same preprocessing step as above to the seed raw dataset when generating the synthetic datasets. We kept the privacy game's attacker training unchanged as in SDR, but in the evaluation phase, we accounted for the non-member outlier to mimic the real challenge faced by an attacker and to eliminate the limitation discussed in \ref{subsec_flawinGame} using the modified game in Fig. \ref{fig:mia_game}.

\subsection{Measuring Utility}
We measured the utility of the derived datasets from two perspectives: (i) their performance on a machine-learning task and (ii) their statistical resemblance to their seed raw data. We used the same task of predicting the \texttt{RiskMortality} attribute to measure the machine learning performance. However, although classification accuracy provides valuable insights into the performance of a specific task, they may overlook other important characteristics and nuances present in the data. To address this concern, we employed statistical metrics, including the Kolmogorov-Smirnov (KS) test \cite{baowaly2019synthesizing} and total variation distance (TVD) \cite{tao2021benchmarking}, to compare the distribution of each column in the derived dataset with its corresponding column in the seed dataset. We then reported the aggregated score, similar to the approach used in Synthcity and Synthetic Data Vault \cite{SDV,Schaar2023Synth}.

\subsubsection{Machine learning utility} \label{subsub_MLclassify}
Like SDR, we used a Random Forest classifier with 100 estimators (with Gini impurity as the splitting criterion) to predict the attribute \texttt{RiskMortality}. We evaluated the performance on 100 synthetic datasets of 1,000 records for each configuration (e.g., for each $\epsilon$ value) of synthetic data generation. We used the average accuracy score on these 100 datasets as the utility metric. Similarly, we averaged the accuracy score of 100 anonymized datasets.

To set up this experiment, we used the 100K-record Texas Hospital Discharge dataset (from the GitHub repository of SDR \cite{stadler2022synthetic}). We removed the target outliers from this population. We capped the outliers to 95\% quantile. Then, we randomly selected 10K records (to be used for the 1K-record dataset samples, which act as the source/seed raw dataset to be anonymized/synthesized). To make the raw test set disjoint, we randomly sampled 5K records (test set) from the rest of the 90K records. For member sets, we added the capped target outlier to the 1K-record raw dataset samples. We report the average classifier accuracy for 500 synthetic datasets (5 outliers, 10 instances of generators, 10 dataset samples per generator).

\subsubsection{Statistical distribution measure} \label{subsub_statistical}
We compared the distributions of the anonymized or synthetic dataset to their corresponding 1K-records raw seed datasets to estimate the quality. We used two robust statistical tests: the Kolmogorov-Smirnov (KS) test \cite{hernandez2022synthetic,baowaly2019synthesizing,SDV,hodges1958kstSig} for the numerical columns and the Total Variation Distance (TVD)\cite{tao2021benchmarking,SDV} for the categorical columns. A brief description of the two metrics follows.

\textbf{Two-sample KS-test complement.} This test is a powerful statistical method used to assess whether two independent samples originate from the same continuous distribution (i.e., same populations) \cite{hodges1958kstSig}. It is a non-parametric approach and does not rely on any specific distribution assumptions, making it suitable for various types of data. With two observed cumulative distribution functions $F(x)$ and $G(x)$ of sizes $n$ and $m$, it is formulated as:
$$KS_{n,m} = \max_x \left| F_n(x) - G_m(x) \right|.$$
We utilized the KSComplement metric ($1 - \textit{KS}$ statistic) provided in Synthcity \cite{Schaar2023Synth}, which generates a score between 0.0 and 1.0. A score of 1.0 indicates that the real data is identical to the synthetic data.

\textbf{Total variation distance (TVD) complement.} This is a non-parametric statistic as well. The TVD measures the largest possible difference between two probability distributions. To compute this metric, at first, the frequency of each category value is determined and expressed as probabilities. Then, the TVD statistic measures and sums up the differences in these probabilities, as demonstrated in the formula. 
$$ TVD(R, S) = \frac{1}{2}\sum_{c \in cat} |(R_{c} - S_{c})|.$$

In this formula, ‘cat’ represents all the categories within a column while ‘R’ and ‘S’ respectively symbolize the probabilities from the raw and the sanitized/synthetic column under consideration. The TVD-complement is calculated as (1 - TVD). The metric ranges from 0.0 to 1.0, with 1.0 indicating identical distribution and 0.0 signifying maximum dissimilarity \cite{SDV,TVcomplement}.

\subsection{Measuring Privacy using MIA Attack}\label{subsecOurMia}
Our adaptation of the MIA attacks focus on machine learning models trained on synthetic or sanitized data, aiming to determine whether a specific individual’s data was utilized in the training of the model. For a model trained with synthetic data, the threat model is described using the following privacy game.
Let $\mathcal{D}_R$ be an underlying real data distribution.

\begin{enumerate}
    \item The challenger samples two disjoint raw datasets: $
    R_{\text{train}}, R_{\text{non-train}} \sim D_ \mathcal{R}^{n} \quad (R_{\text{train}} \cap R_{\text{non-train}} = \emptyset). $ 
    
    \item The challenger trains a synthetic data generator using $R_{\text{train}}$ and a generates a synthetic dataset S. Then, using the dataset S, a machine learning classifier ML is trained. 
    
    \item The challenger chooses a bit \textit{b} uniformly at random from \{0, 1\} and samples a record $\mathcal{X}$. If $\textit{b=0}$, the challenger samples $\mathcal{X}$ from $R_{\text{non-train}}$ or, if $\textit{b=1}$, samples $\mathcal{X}$ from $R_{\text{train}}$. The challenger gives $\mathcal{X}$ to the adversary.
    
    \item The adversary, with access to $\mathcal{D}_R$ and query access to the model ML, predicts \^{b}. If \textit{\^{b}=b}, the adversary wins. Otherwise, the adversary loses.
\end{enumerate}

The attacker's advantage (AA) is defined as: 
$$AA = \Pr[\hat{b} = 1 \mid b = 1] - \Pr[\hat{b} = 1 \mid b = 0]$$
$$\Rightarrow \text{True Positive Rate} - \text{False Positive Rate}$$

Note that the adapted MIA attack on the machine learning model does not directly target the records in the victim model’s training dataset (since it is synthetic data; ref. Algorithm 1). Instead, it attempts to launch an indirect attack against the records in the seed dataset used to generate the training data for the victim model. The attacker assumes that the victim model, trained on synthetic data, may behave differently when presented with seed data from the synthetic data generator compared to non-seed data.

\textbf{Note on MIA attack against \textit{k}-anonymity.} For \textit{k}-anonymized data, the MIA attack is mounted against a victim model trained on a \textit{k}-anonymized dataset (following the game in Section \ref{subPriv}). The attacker's objective is to infer whether a target record was used to train the victim classifier. Given that the victim model remains the same, these two MIA attacks (i.e., on models (1) trained indirectly through synthetic datasets and (2) trained directly on \textit{k}-anonymized datasets) provide a way to compare how various privacy preservation techniques perform in protecting the privacy of participating records after the release of a downstream model (i.e., the victim model).

In the case of dataset release  (as opposed to model release), \textit{k}-anonymized data is directly vulnerable to membership inference attacks, especially when the adversary has background knowledge of the underlying data distribution $D_{\mathcal{R}}^{n}$. If the entire population is included and a \textit{k}-anonymized dataset is released, the adversary can infer with 100\% accuracy whether an individual's record is part of the released dataset by matching the quasi-identifiers of the target against the quasi-identifier set in the released dataset. However, if only a portion of the population is included, then such direct quasi-identifier comparison will yield a membership accuracy of $$\frac{| \text{QID}_{\text{target}} |_{\text{published}}}{| \text{QID}_{\text{target}} |_{\text{population}}}.$$ In contrast, such a direct comparison attack is not effective against synthetic data.

\textbf{MIA attack implementation.} Now, we explain how we implemented the membership inference attack using the TensorFlow Privacy library’s membership inference attack (released by a team from Google Research \cite{tf2020tfmia}, used in many researches (e.g., \cite{tang2022mitigating,song2021systematic})). Its MIA attack is inspired by the original attack proposed by Shokri et al. \cite {shokri2017membership} but has a slightly different approach. The attack does not involve training multiple shadow models as described in the original attack. Instead, it leverages the findings from \cite{salem2018ml}, which indicate that one shadow model could be adequate. The Tensorflow-library's MIA attack uses the original model’s predictions on the target data points to determine their susceptibility to membership inference. This library launches the MIA attack from the perspective of the owner of a (victim) model, who wants to assess the model's MIA privacy state. To capture the worst-case leakage, a shadow model needed to behave the same as the target model. The TensorFlow-privacy library’s MIA attack does this exact replication of behavior by making disjoint subsets of the victim model’s labeled prediction vectors and uses these disjoint subsets to act as both the shadow dataset and the target dataset (but not simultaneously, i.e., when one subset acts as the shadow dataset, another subset acts as the target dataset). This approach eliminates the need to train shadow models that approximate the original model’s behavior, making the attack more efficient. Essentially, the original model being targeted acts as its own shadow model, perfectly approximating its behavior. Algorithm \ref{algo:membership_attack} summarizes the process.
\setlength{\textfloatsep}{12pt} 
\begin{algorithm}[t]\small
\itemsep5pt
\caption{MIA Attack using TensorFlow-Privacy} \label{algo:membership_attack}
\begin{algorithmic}

    \State\textbf{Synthetic data generation and victim model training}\\
\vspace{2pt}

    \graycomment{\#Sample raw datasets}
    \State {1. }$R_{train},R_{non-train} \sim D_ \mathcal{R}^{n}; (R_{non-train} \cap R_{train}=\emptyset)$\\
\vspace{2pt}
    \graycomment{\#Train a generative model}
    \State {2. }$g(R_{train}) \sim GM(R_{train})$\\
\vspace{2pt}
    \graycomment{\#Sample Synthetic data}
    \State {3. }$S \sim D^m_g(R_{train})$\\
\vspace{2pt}
    \graycomment{\#Train victim Classifier Model}
    \State {4. }$cv(S) \sim CM(S)$
    \vspace{6pt}
\State \textbf{Attacker Training \& Testing using cross validation}\\
\graycomment{\# Get IN \& Out prediction vectors}
\State {5. }$Pred_{in} \leftarrow cv(R_{train})$\\
\hspace{2 mm}$Pred_{out} \leftarrow cv(R_{non-train})$\\
\vspace{2pt}

\graycomment{\# Divide prediction vectors into equal folds}
\State {6. } Define the number of folds (k) for cross-validation and\\
\hspace{3.5 mm} split the $Pred_{in}$ and $Pred_{out}$ dataset into k equal-sized folds.\\
\vspace{4pt}

\graycomment{\# Train the MIA model m and gather the attack results}
\State {7. } \textbf{For each} fold i from 1 \textbf{to} k \textbf{do}:\\
\hspace{7 mm}  i. Use fold i from $Pred_{in}$ and $Pred_{out}$ as the test set.\\
\hspace{7 mm} ii. Combine the remaining folds into the training set.\\
\hspace{7 mm}iii. Train the attack model m using the training set.\\
\hspace{7 mm}iv. Evaluate the model's performance on the test set.\\
\vspace{4pt}

\graycomment{\# Aggregate the results.}
\State {8.  }{Compute the average performance metric across all folds.}
\end{algorithmic}
\end{algorithm}

Here $R$ is the population of real data records. Initially, two sets of datasets are sampled: $R_{train}$ and $R_{non-train}$, each drawn from the real data distribution $D_R^{n}$.
The MIA attack on synthetic datasets involves two classifiers: the victim classifier \textit{cv} (here, the \texttt{RiskMortality} classifier) and the membership-inference classifier, \textit{m}. 
In this scenario, there is a synthetic data generator $g(R_{train})$ (or a sanitizer) that takes real data as input and produces synthetic data (or sanitized data) $S$ which is used to train the victim classifier, $cv$.

\textbf{Attacker training and evaluation.} The owner uses the victim model to derive \texttt{RiskMortality} prediction vectors for the training dataset (\textit{$R_{train}$} for synthetic, or $S$ for \textit{k}-anonymized; label: member) and non-training dataset \textit{$R_{non-train}$} (label: non-member). The owner then divides the prediction vectors into equal folds.
Using these folds, the owner then trains the attack models, i.e., the membership inference classifiers. Finally, the trained MIA attackers are tested using the other folds (i.e., excludes the training fold) and the attacker's membership predictions are compared against the ground-truth labels.

Additionally, we also used SDR's privacy game but accounted for the non-member outlier in the evaluation phase (Fig. \ref{fig:mia_game}). To remove the bias discussed in subsection \ref{subsec_unfair}, we capped the outliers to 95\% quantile.
\begin{table}[t]
\centering
\small  
\begin{tabular}{m{1.2cm} m{1.0cm} p{0.75cm} p{0.6cm} p{0.5cm} c} 
\multicolumn{2}{l}{\multirow{2}{*}{\centering\textbf{}}} & \multicolumn{2}{c}{\textbf{Utility}} & \multicolumn{2}{c}{\textbf{Privacy}} \\
\cmidrule(lr){3-4} \cmidrule(lr){5-6}
\multicolumn{1}{l}{\textbf{Method}} & \multicolumn{1}{l}{Params.} & Stat. & ML  & AA & OC \\ 
\midrule

\multirow{4}{1.5cm}{NHS} & \textit{k}=5 & 0.97 & 0.68 & 0.81 & 4 \\  

 & \textit{k}=10 & 0.95 & 0.65 & 0.78 & 3\\  
 & \textit{k}=15 & 0.94 & 0.65 & 0.76  & 3\\  
 & \textit{k}=20 & 0.90 & 0.64 & 0.73  & 2\\ \hline  
\multirow{4}{1.5cm}{BayNet} & Bin=2 & 0.92 & 0.72 & 0.65 & 0 \\ 

 & Bin=5 & 0.97 & 0.72 & 0.67  & 0\\  
 & Bin=10 & 0.97 & 0.72 & 0.70  & 0\\ 
 & Bin=25 & 0.97 & 0.72 & 0.71 & 0 \\ \hline  
\multirow{4}{1.5cm}{PrivBayes} & $\epsilon$=0.1 & 0.60 & 0.23 & 0.75  & 0\\ 
 & $\epsilon$=1.0 & 0.67 & 0.46 & 0.76 & 0 \\  
 & $\epsilon$=10.0 & 0.82 & 0.65 & 0.81 & 0 \\ 
 & $\epsilon$=15.0 & 0.85 & 0.66 & 0.81  & 0\\ \hline 

 \multirow{4}{1.5cm}{PATEGAN} & $\epsilon$=0.1 & 0.71 & 0.51 & 0.52   & 0\\ 
 & $\epsilon$=1.0 & 0.80 & 0.59  & 0.63 & 0 \\  
 & $\epsilon$=10.0 & 0.88 & 0.63 & 0.67 & 0 \\ 
 & $\epsilon$=15.0 & 0.88 & 0.63 & 0.67  & 0\\ \hline 

 \multirow{4}{1.5cm}{DDPM} & ts=100 & 0.73 & 0.63 & 0.52  & 0\\ 
 & ts=500 & 0.71 & 0.63 & 0.50 & 0 \\  
 & ts=1000 & 0.70 & 0.63 & 0.51 & 0 \\ 
 & ts=1500 & 0.70 & 0.63 & 0.51  & 0\\ \hline 

\end{tabular}
 \vspace{0.5cm}
\caption{Results on Texas: We varied parameter \textit{k} for \textit{k}-anonymization, bin for Bayesian Network, $\epsilon$ for PrivBayes and PATE-GAN, and timesteps for DDPM. For utility, we report the statistical similarity score (stat.) \& the machine learning accuracy (ML). For privacy, we took attacker advantage (AA) from MIA attack, and the outlier-count (OC) indicates the number of outliers detected out of the 5 outliers with a precision rate as low as 60\% using the SDR's game with non-member outliers considered in the evaluation.}
\label{tab:my_table1}
\end{table}

\section{Results}\label{secResults}

Table \ref{tab:my_table1} displays a summary of our findings concerning various data sanitization techniques applied to the Texas dataset. The ‘ML’ column in the table represents the accuracy of the classification where \texttt{RiskMortality} was the target column. It also shows the attacker's advantage (AA) from the TensorFlow-MIA attack. In the rightmost column, the outlier count (OC) indicates the number of outliers detected out of the five outliers, with a precision rate as low as 60\% using the SDR game, accounting for the case of non-member outliers in the test dataset. (We experimented with higher precision rates; if higher precision rates are considered, none of the outliers are detected.)
\begin{figure*}[t]
  \centering
  \includegraphics[width=1\linewidth]{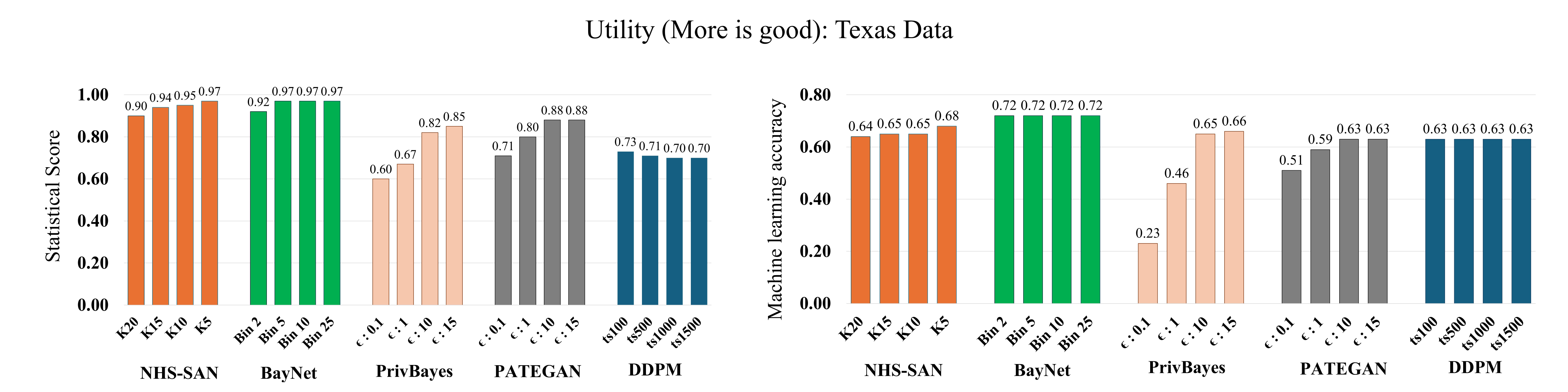}
  \caption{Utility of the datasets produced by k-anonymization and various synthetic data generators. (Left bar graph: statistical similarity score; right graph: classification accuracy when the classifier is trained on the resultant data)}
  \label{fig:Utility_BN_NHS}
\end{figure*}

\subsection{Utility}
Fig. \ref{fig:Utility_BN_NHS} presents a utility comparison between \textit{k}-anonymized datasets and various synthetic datasets. The left bar graph shows the aggregate statistical score for resemblance. The observed trend is consistent. The aggregate score for BayNet with bin=2 is 92\%, which reaches 97\% when bin is increased to 5. In the case of the \textit{k}-anonymized dataset, the trend is similar. For example, with \textit{k} set to 5, the score is 97\%, but when \textit{k} is increased to 20, the score drops to 90\%. 
The right bar-graph displays the machine learning utility of the datasets using the classification accuracy score. For synthetic data generated using BayNet, the classification score remains consistent at 72\%. However, the score decreases significantly with the increasing values of \textit{k}. For \textit{k} = 5, the classification score is relatively high at 68\%. As the value of \textit{k} increases, classification accuracy decreases. For \textit{k} = 20, the classification accuracy dropped to 64\%, which is close to the majority class classification score. This decline can be attributed to the suppression-based nature of the NHS-sanitization, which reduces the number of records in the anonymized dataset as \textit{k} increases. The utility and privacy of PrivBayes and PATE-GAN vary with different values of epsilon, following the expected direction of change. DDPM’s utility and privacy fluctuate slightly with respect to timesteps but remain stable. These trends in utility are also consistent across datasets, as evidenced by the Adult dataset and Credit card dataset (Figures \ref{fig:Utility_adult} and \ref{fig:Utility_credit} in Appendix C).
\begin{figure}[t]
  \centering
  \includegraphics[width=1\linewidth]{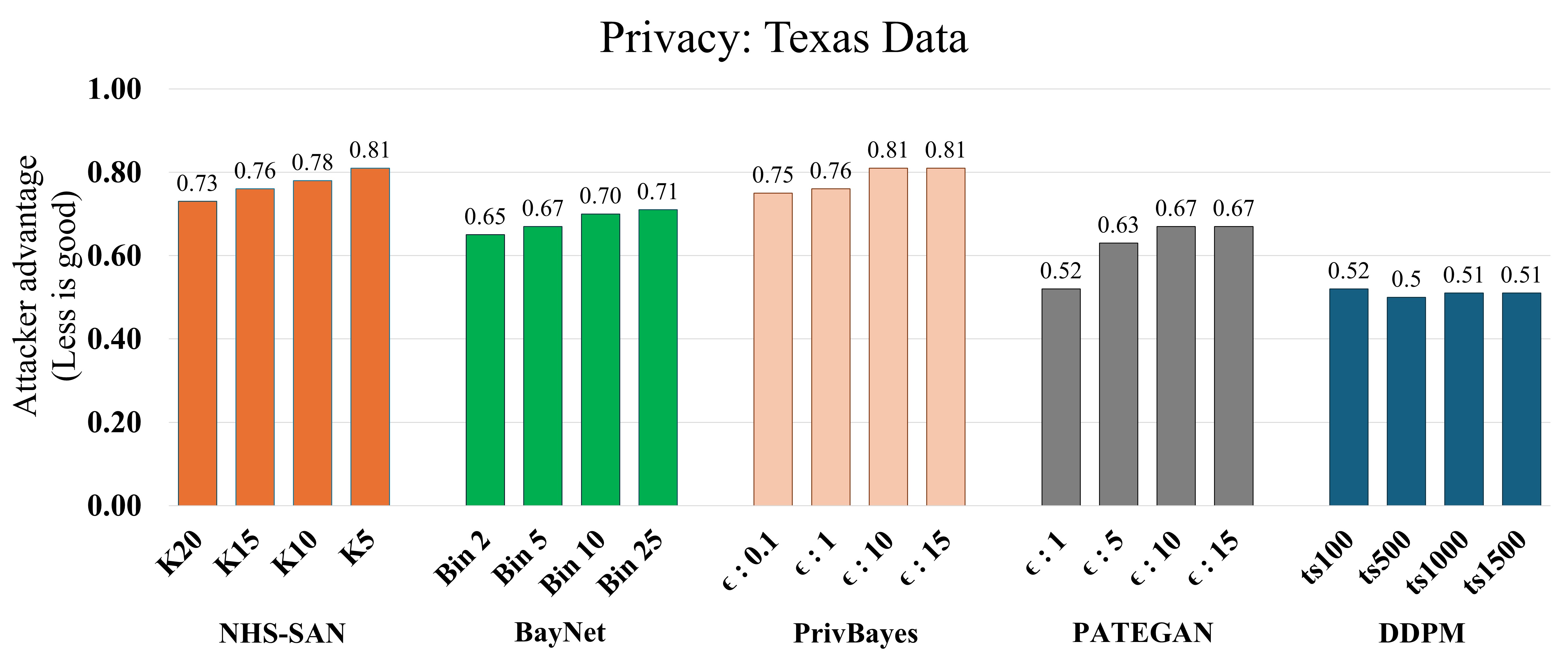}
  \caption{Attacker's advantage using the MIA attack. It demonstrates the changing of privacy levels for different settings of the \textit{k}-anonymization and synthetic data generators. (Dataset: Texas) }\label{fig:Privacy_MIA_BN_NHS}
\end{figure}

\begin{figure*}[htb]
  \centering
  \includegraphics[width=1\linewidth]{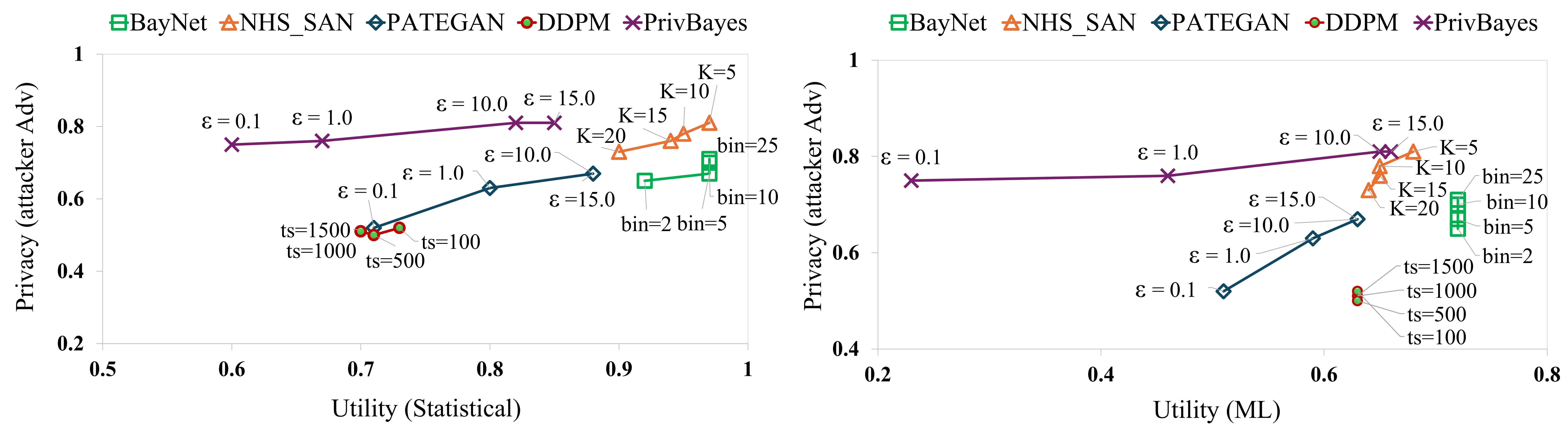}
  \caption{Privacy-Utility Trade-off on Texas dataset. Models: BayNet (BN), \textit{k}-anonymization (NHS\_SAN), PATEGAN, DDPM, PrivBayes. Hyper-parameter: BN (bin: 2, 5, 10, 25), NHS\_SAN (\textit{k}: 5, 10, 15, 20), PATEGAN and Privbayes ($\epsilon$: 0.1, 1.0, 10.0, 15.0), and DDPM (diffusion timesteps, ts: 100, 500, 1000, and 1500).}
  \label{fig:GHpriv_vs_util_AttAdv}
\end{figure*}
\subsection{Privacy}
Figure \ref{fig:Privacy_MIA_BN_NHS} here and Figures \ref{fig:Privacy_adult} and \ref{fig:Privacy_credit} in the Appendix illustrate the results of the TensorFlow-membership attacker's advantage on data derived from the Texas, Adult, and Credit card datasets. We note the following: 
\begin{itemize}
    \item The success of MIA attacks can vary significantly across different datasets, even when the same privacy parameter (e.g., \textit{k} in \textit{k}-anonymization, or $\epsilon$ for a differentially private model) is used. For example, an attacker had average attack advantages of 0.67, 0.55, and 0.04 on models trained with PATE-GAN ($\epsilon=5$)-generated data from the Texas, Adult, and Credit Card datasets, respectively. Therefore, a parameter that ensures privacy at a certain level for one dataset may not achieve the same privacy level for another dataset. Additionally, some datasets may be inherently more resistant to MIA attacks (e.g., attacker advantage of 0.14 for the Credit Card dataset compared to 0.81 for the Texas dataset).
    
    \item The relative privacy performance of synthetic/anonymization models may vary across different datasets. For instance, the position of DDPM compared to other models on the Texas dataset changes when applied to the Adult and Credit Card datasets. This suggests that a synthetic data generator that outperforms a particular model on one dataset may not necessarily do so on another. 
    
    \item Interestingly, while \textit{k}-anonymity shows a clear trend in terms of utility (i.e., utility decreases as \textit{k} increases), the same cannot be said for privacy. Although the expected trend is observed with the Texas dataset (privacy increases as \textit{k} increases), this does not hold for the Adult and Credit datasets. In both the Adult and Credit datasets, \textit{k}=5 provides better privacy than \textit{k}=10 or higher. A similar phenomenon is observed with BayNet and DDPM, though in a smaller scale (e.g., DDPM ranges from 0.50 to 0.52). Notably, PATE-GAN consistently displayed the expected trend in attacker's advantage relative to $\epsilon$.
    
\end{itemize}

The unpredictable behavior of \textit{k}-anonymization regarding membership inference attacks across some datasets raises new research questions. Several potential hypotheses could explain this apparent fluctuation. It's important to note that \textit{k}-anonymity only focuses on the quasi-identifier attributes. As Carlini et al. \cite{carlini2022membership} suggested, some data points are inherently harder to fit than others. Take, for instance, the Credit Card data (Fig. \ref{fig:Privacy_credit}). It is possible that, as \textit{k} increases from 5 to 15, hard-to-fit cases remain while other records get removed. With fewer records at \textit{k}=15 than at \textit{k}=5, if the remaining cases are hard-to fit and the model is overfitted, the attacker's advantage could increase, as the model may focus more on those hard-to-fit records. Conversely, at \textit{k}=20, these hard-to-fit records might get removed, leading to a sudden decrease in the attacker's advantage. Another hypothesis is that when the training dataset is large, the impact of a single record becomes less significant. Consequently, removing records through \textit{k}-anonymization may reduce the training dataset size and lead to overfitting in the model increasing the attacker advantage. Further research is needed to explore this phenomenon and the hypotheses.

From the rightmost column in table \ref{tab:my_table1}, we can see that SDR's privacy game was not successful for synthetic dataset up to the attacker's precision of 0.6. With the equal number of member and non-member test cases, a precision of 0.6 is a very low bar for the attacker \cite{carlini2022membership}. Moreover, for NHS-sanitization, SDR's privacy game was unable to detect the membership of any outlier up to a precision of 0.8. When we lowered the attacker's precision bar further to 0.6, it was able to detect some of the outliers. Note that Carlini et al. \cite{carlini2022membership} suggests a false positive rate below 0.1. An attacker's precision of 0.6 or 0.8 is not considered a powerful attack, as it still provides significant grounds for deniability.

\subsection{Trade-off Visualization}\label{reportingTemplate}
Using the separate privacy and utility graphs presented so far, it is difficult to decide which algorithm is better for a particular level of privacy and utility. We now present a better way to depict the privacy-utility trade-off of different algorithms on a single graph. Fig. \ref{fig:GHpriv_vs_util_AttAdv} provides the visual representations of the privacy-utility trade-offs for both synthetic data and \textit{k}-anonymized data. This visualization approach simplifies the problem of finding which method performs better in terms of both privacy and utility. It is well-established in the literature that increasing privacy may lead to a loss of utility. Using this visualization, a practitioner/researcher can easily choose the appropriate algorithm and the right parameter that satisfies the intended level of privacy and utility.

The graphs in Fig. \ref{fig:GHpriv_vs_util_AttAdv} presents the privacy-utility trade-off on Texas dataset. The graph on the left shows the trade-off in terms of attacker advantages (MIA) and statistical utility. The graph on the right depicts the privacy-utility trade-off in terms of attacker advantages (MIA) and machine learning performance. Figures \ref{fig:Adult_tradeoff} and \ref{fig:credit_tradeoff} in Appendix C present the  trade-off graphs for Adult and Credit card datasets. 

Some of the models show nearly horizontal or vertical trends in the privacy-utility trade-off figures for different values of the parameters. While the parameters are expected to influence the property of the generated data, these do not necessarily impact the core information required for prediction tasks. Machine learning models often rely on high-level patterns, which remain unaffected by minor statistical changes caused by different parameter settings. 
For example, in Fig. \ref{fig:GHpriv_vs_util_AttAdv}, we observe that model accuracy saturates in the case of DDPM and BayNet once a baseline level of performance is reached. In Fig. \ref{fig:credit_tradeoff} in Appendix C, we observe that the attack was barely successful against synthetic data. The advantage (TPR - FPR) is very small, \~ 0.05, indicating that the privacy guarantee remains very strong (even for larger privacy budget such as $\epsilon$=15). Consequently, the privacy guarantee remains robust, resulting in the observed horizontal trend.
  
Regulatory agencies, such as the European Medicines Agency and Health Canada, require a re-identification risk threshold of 0.09, i.e., $k \geq 11$ for \textit{k}-anonymization \cite{maritsch2022data,branson2020evaluating}. In the case of the Texas dataset, since the trade-off curve for BayNet consistently lies below that of the \textit{k}-anonymized data for $k \geq 10$, it is evident that BayNet provides better utility and privacy for all utility levels achievable by synthetic data. For instance, if the statistical utility requirement is 0.96 or lower, synthetic data is always preferable. However, at a higher utility level (e.g., 0.98), which may not yet be attainable by synthetic data, practitioners must exercise caution as the attacker's advantage may increase. At reasonable utility levels, PrivBayes offers less protection against MIA attacks compared to other anonymization techniques. While DDPM and PATE-GAN provide better privacy than \textit{k}-anonymization, they fall short in terms of utility at least till $k=20$. In case of the Adult and Credit Card datasets, BayNet continue to show better performance when $k \geq 10$. Interestingly, in Adult dataset, PATE-GAN also offered better trade-off than \textit{k}-anonymization when $k \geq 15$ (Fig. \ref{fig:Adult_tradeoff}).

From these three experiments, we may conclude that not all synthetic data generation models provide inherently superior trade-offs compared to \textit{k}-anonymization across all datasets and trade-off levels. However, in each experiment, we identified at least one synthetic data generator that offers a better privacy-utility trade-off than \textit{k}-anonymization while still complying with regulatory requirements.

\section{Conclusion}\label{secConclusion}

Until recently, synthetic data was considered to offer better privacy-utility trade-offs compared to traditional anonymization techniques. However, a recent article has contradicted the longstanding position that synthetic data is better than traditionally sanitized data for preserving utility while maintaining privacy. In this article, we examined the implementation of the novel privacy game and identified critical characteristics that were limiting the scope of their findings. We also provided general but comprehensive recommendations to avoid potential pitfalls in future research endeavors. Then, we experimented with synthetic data and \textit{k}-anonymization, evaluating their performance. To make our evaluation fair among the competing techniques, the input dataset samples received the same preprocessing treatment. We compared the privacy-utility trade-off using established statistical metrics and membership inference attacks. We incorporated a visual reporting template for illustrating the quantitative interpretations of the privacy-utility trade-off. This visualization method offers a more informative view and helps the data publishers select the data publishing techniques that align with their desired level of trade-off. Our findings showed a generally predictable trend in the trade-off dynamics across different privacy and utility requirements levels. We also found that not all synthetic data generators provide a better privacy-utility tradeoff, but certain synthetic data approaches outperform the \textit{k}-anonymization method.

\textbf{Future work.} In this study, we did not explore the broader question of whether some synthetic data generation model is always better than \textit{all} traditional anonymization techniques because it involves a much bigger scope. Traditional anonymization methods encompass a variety of techniques and extend beyond simple data suppression. These techniques incorporate a range of generalization-based algorithms, such as the Datafly, Mondrian, and Incognito, to name a few. We also limited our analysis to basic statistical utility metrics even though there are more utility metrics that consider different aspects of the data. Some of these metrics do not work well with generalized anonymized data. Furthermore, there are many variations of membership inference attack \cite{tang2022mitigating} and also other attacks beyond membership inference attacks. In the future, we plan to conduct a more comprehensive study that includes prominent anonymization algorithms, a wider range of utility metrics, and various privacy attacks to thoroughly assess whether current synthetic data generation techniques provide a more effective solution than the best methods from traditional anonymization approaches.

\begin{acknowledgements}
 A.R.S. is a Gordon P. Osler scholar and was partially supported by UMGF fellowship. N.M. was supported by the NSERC Discovery Grants (RGPIN-04127-2022) and NSERC Alliance Grants (ALLRP 592951 - 24).
\end{acknowledgements}
\section*{Data availibility}
The datasets are publicly available: Texas hospital discharge \cite{ground2021git}, Adult income\cite{adult2}, Credit card churn \cite{CreditCardChurn}.
\section*{Declarations}
\begin{itemize}
    \item{Conflict of interest/Competing interests}
    \begin{itemize}
            \item[] The authors have no conflicts of interest to declare.
    \end{itemize}

    \item Authors' contributions
    \begin{itemize}
        \item[] All authors participated in the design of the methods. F.J.S. implemented the methods in python and conducted the experiments. All authors wrote, reviewed and revised the paper. Y.W. and N.M. supervised the research.
    \end{itemize}

\end{itemize}

\section*{Appendix A: Abbreviations}
Table \ref{tab:abbreviation} outlines the abbreviations used throughout this paper.

\begin{table} [H]
\centering
\begin{tabular}{ m{1.8cm}  m{6.0cm} }
\hline
\textbf{Abbreviation}  & \textbf{Description} \\
\hline
AA & Attacker advantage \\
AUC & Area under the curve \\
BayNet & Bayesian networks \\
CTGAN & Conditional Tabular GAN \\
DPGAN & Differentially Private GAN \\
DDPM & Denoising diffusion probabilistic models\\
FPR & False positive rate  \\
GANs & Generative adversarial networks \\
GDPR & General Data Protection Regulation \\
GM & Generative model  \\
HIPAA & Health Insurance Portability and Accountability Act  \\
KL & Kullback–Leibler divergence  \\
KS & Kolmogorov-Smirnov test \\
MIA & Membership inference attack \\
ML & Machine learning accuracy \\
NHS & National Health Service, England  \\
NIST & National Institute of Standards and Technology  \\
PATE-GAN & Private Aggregation of Teacher Ensembles GAN \\
PG & Privacy Gain \\
PPDP & Privacy-preserving data publishing \\
PrivBayes & Privacy-Preserving Bayesian Network \\
QIDs & Quasi-identifiers \\
SAN & NHS sanitization  \\
SDG & Synthetic data generation  \\
SDR & Synthetic data release  \\
TabDDPM & DDPM for tabular data\\
TPR & True positive rate \\
TVD & Total variation distance \\
VAEs & Variational auto-encoders \\
\hline

\end{tabular}
\caption{List of abbreviations}
\label{tab:abbreviation}
\end{table}

\section*{Appendix B: Datasets and Generative Models}

\texttt{Texas hospital discharge dataset \cite{stadler2022synthetic}:} It contains 18 columns and 100,000 records. The patient's state, sex code, race, ethnicity, and age were considered quasi-identifiers. The dataset preparation steps are described in Section \ref{subsub_MLclassify}. The classification problem considered is the same as in SDR: a multiclass classification task on the RiskMortality variable.

\texttt{Adult (census income) dataset \cite{adult2}:} It contains 15 columns and 45,222 records. Age, race, gender, marital status, and native country were considered quasi-identifiers. The same preprocessing steps as in the Texas dataset were applied. The classification problem is the same as in SDR: a binary classification task on the income variable.

\texttt{Credit card customer churn dataset \cite{CreditCardChurn}:} Customers who switch banks are labeled as churners, and the task is to classify them as either churn or non-churn (binary classification problem). The dataset comprises 10127 customers, with 1627 identified as churners, across 20 variables (‘Attrition\_Flag’ as the dependent variable and 19 independent variables). Gender, marital status, dependent count, and total relationship count were considered quasi-identifiers. For preprocessing, the records above the 99th percentile or below the 1st percentile in the numeric columns were removed, resulting in 8921 records. A set of 2000 records was sampled using stratified random sampling on the target variable to be used as the test. To serve as seed sets to anonymizer/synthetic data generators, 100 sets of 1000 records were sampled from the remaining records, also using stratified random sampling.

\textbf{Generative Models}

\texttt{BayesNet}\cite{ping2017datasynthesizer}:
BayesNet creates synthetic datasets by constructing a graphical model that captures the probabilistic relationships among attributes in the original data. It models the joint probability distribution of the data and uses this structure to generate new data points by sampling from the learned distribution. We used the implementation utilized by Stadler et al.\cite{ground2021git}

\texttt{PrivBayes}\cite{zhang2017privbayes}: The algorithm injects noise into the learning of the Bayesian network structure and parameters to satisfy differential privacy. We used the implementation utilized in Synthcity\cite{Schaar2023Synth}.

\texttt{PATE-GAN}\cite{jordon2018pate}: PATE-GAN is a privacy-preserving data generation framework that combines GANs with the PATE (Private Aggregation of Teacher Ensembles) mechanism. It employs multiple teacher models trained on disjoint subsets of the data to guide the training of a student generator model. The PATE mechanism ensures differential privacy by adding noise to the aggregation of teacher outputs, which aims to prevents the leakage of sensitive information. We used the implementation utilized in Synthcity\cite{Schaar2023Synth}.

\texttt{TabDDPM}\cite{kotelnikov2023tabddpm}: TabDDPM (Tabular Diffusion Denoising Probabilistic Model) is a generative model designed for creating synthetic tabular data. It leverages a diffusion process that progressively adds noise to the original data and trains a neural network to reverse this process, reconstructing data samples from noise. We used the implementation utilized in Synthcity\cite{Schaar2023Synth}.

\section*{Appendix C: Results}
For both datasets (Adult and Credit card):
\begin{itemize}
    \item Models: \textit{k}-anonymization (NHS\_SAN), BayNet, PATEGAN, and TabDDPM. 
    \item Hyper-parameter: NHS\_SAN (i.e., \textit{k}-anonymization) (\textit{k}: 5, 10, 15, 20), BayNet (bin: 2, 5, 10, 25), PATEGAN ($\epsilon$: 0.1, 1.0, 10.0, 15.0), and DDPM (diffusion timesteps, ts: 100, 500, 1000, and 1500).
\end{itemize}
\begin{figure}[H]
  \centering  \includegraphics[width=1.0\linewidth]{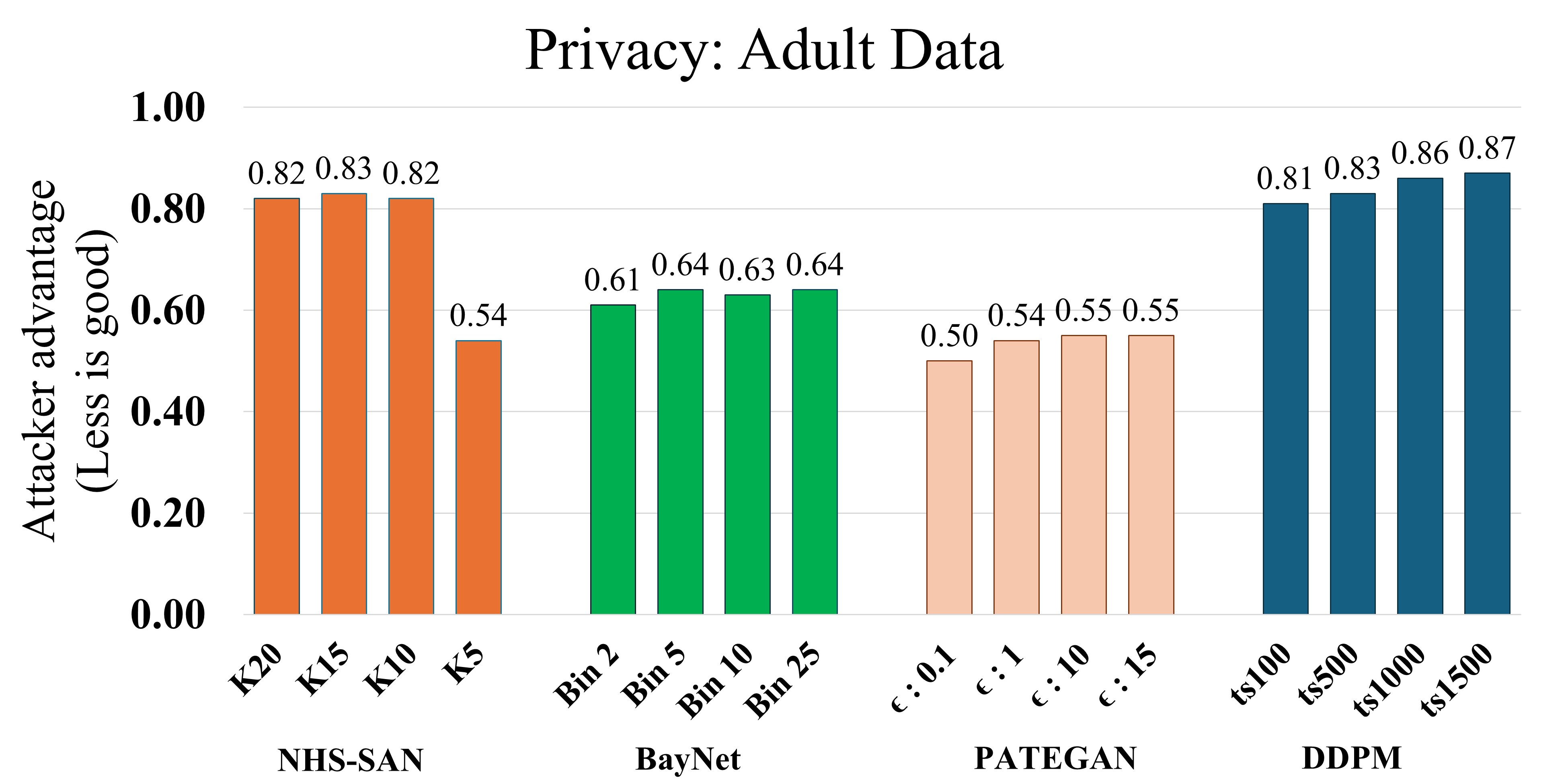}
  \caption{Adult data: Privacy (attacker advantage from MIA).}\label{fig:Privacy_adult}
\end{figure}
\begin{figure}[H]
  \centering  \includegraphics[width=1.0\linewidth]{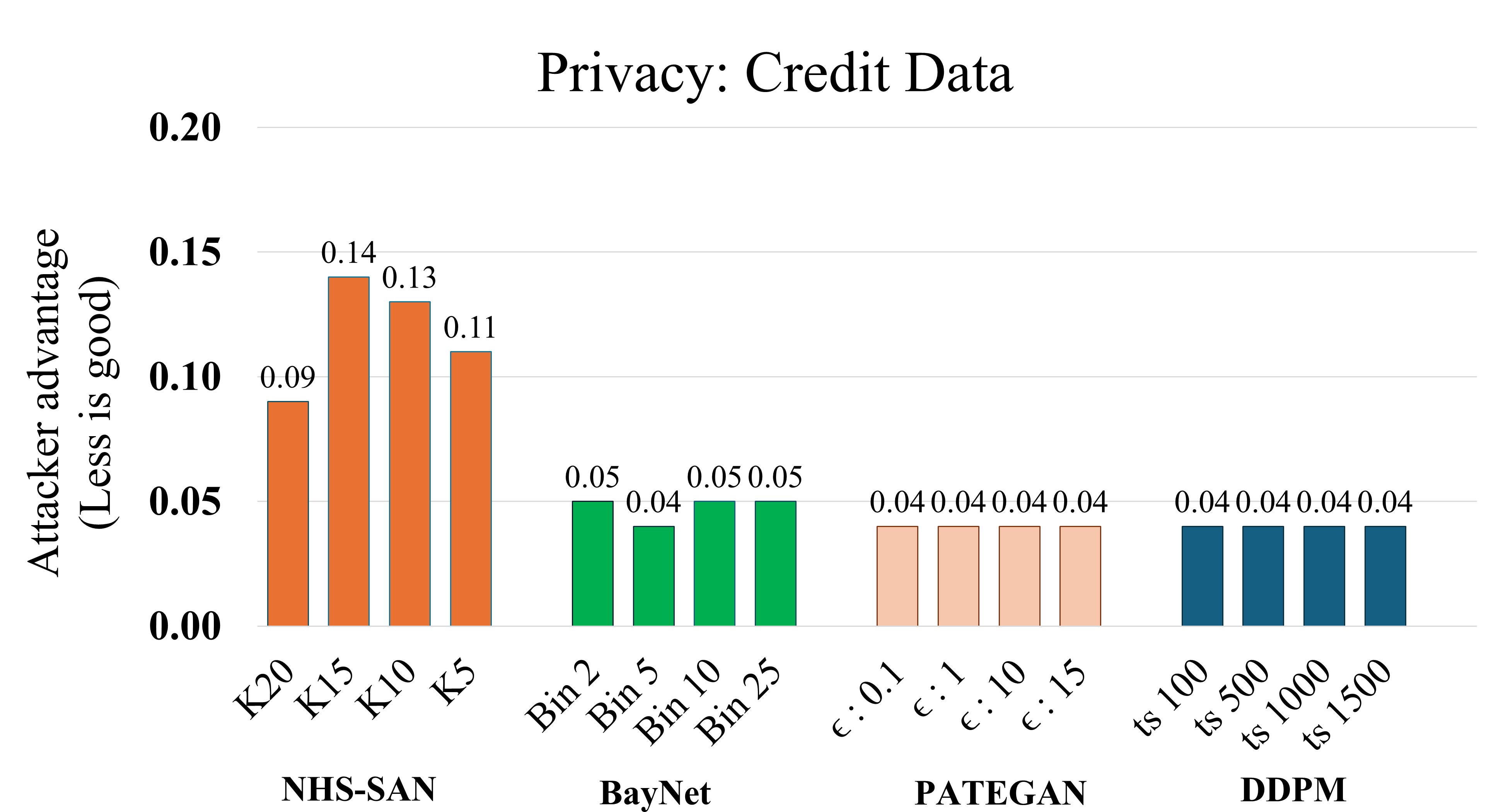}
  \caption{Credit data: Privacy (attacker advantage from MIA).}\label{fig:Privacy_credit}
\end{figure}
\clearpage
\begin{figure*}[htb]
  \centering
  \includegraphics[width=1\linewidth]{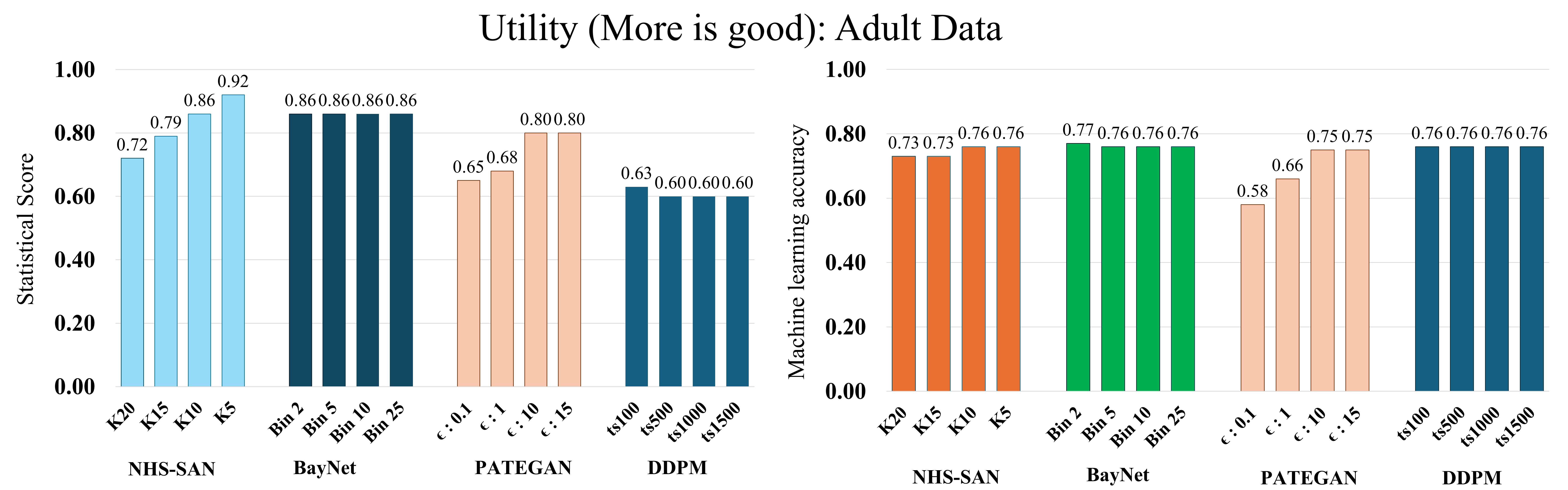}
  \caption{Adult dataset: Utility }
  \label{fig:Utility_adult}
\end{figure*}

\begin{figure*}[htb]
   \vspace{1.5cm}
  \centering
  \includegraphics[width=1\linewidth]{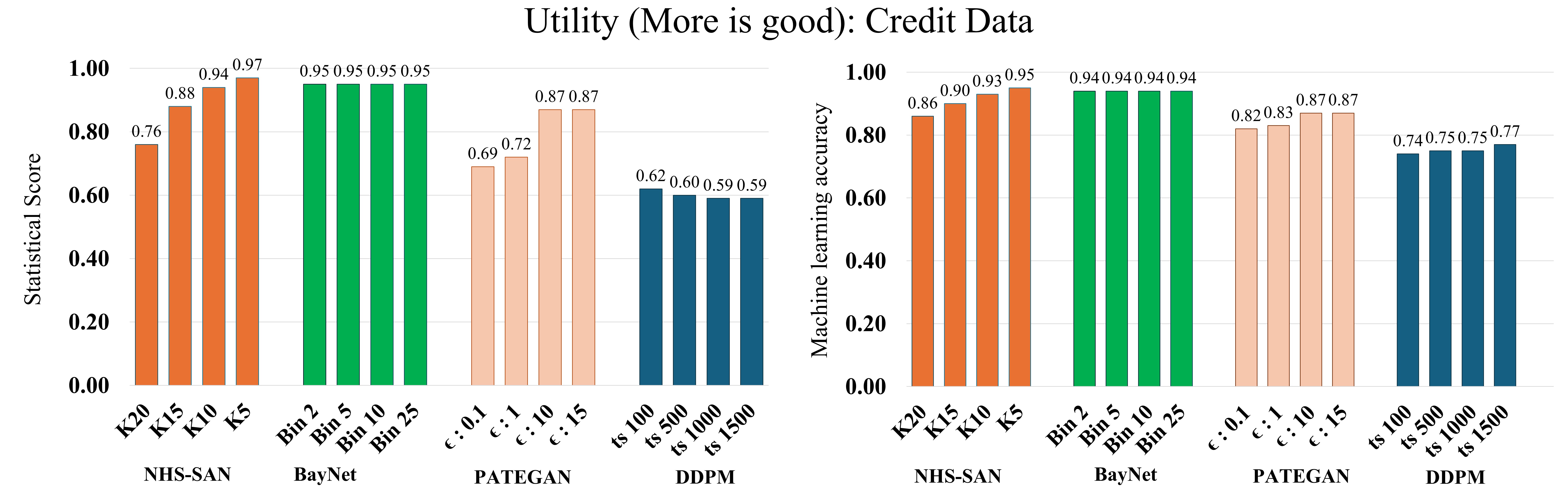}
  \caption{Credit card dataset: Utility}
  \label{fig:Utility_credit}
\end{figure*}
\begin{figure*}[htb]
  \vspace{1.5cm}
  \centering
  \includegraphics[width=1\linewidth]{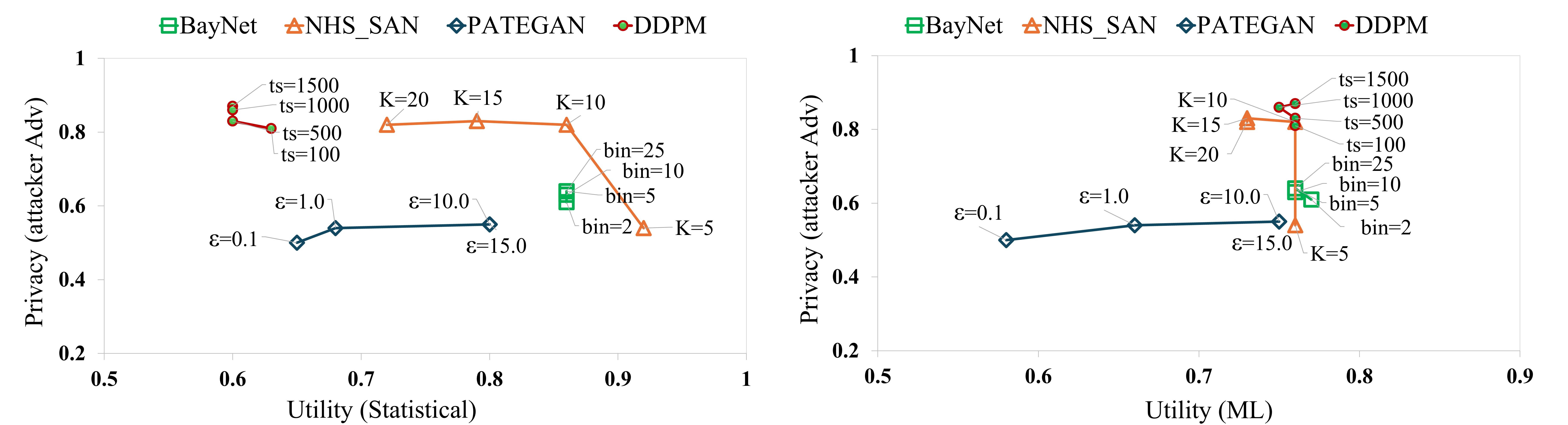}
  \caption{\texttt{Adult} dataset: Privacy-Utility Trade-off.}
  \label{fig:Adult_tradeoff}
\end{figure*}
\FloatBarrier
\clearpage

\begin{figure*}[t]
  \centering
  \includegraphics[width=1\linewidth]{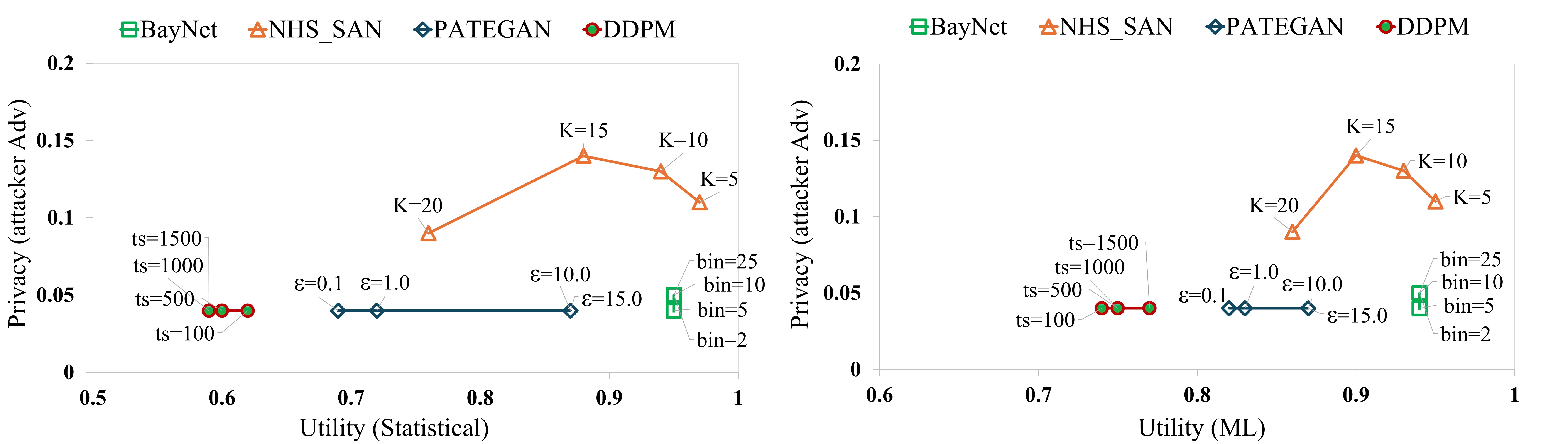}
  \caption{\texttt{Credit card} dataset: Privacy-Utility Trade-off. }
  \label{fig:credit_tradeoff}
\end{figure*}

\begin{figure*}[htb]
  \centering
  \includegraphics[ width=1\linewidth]{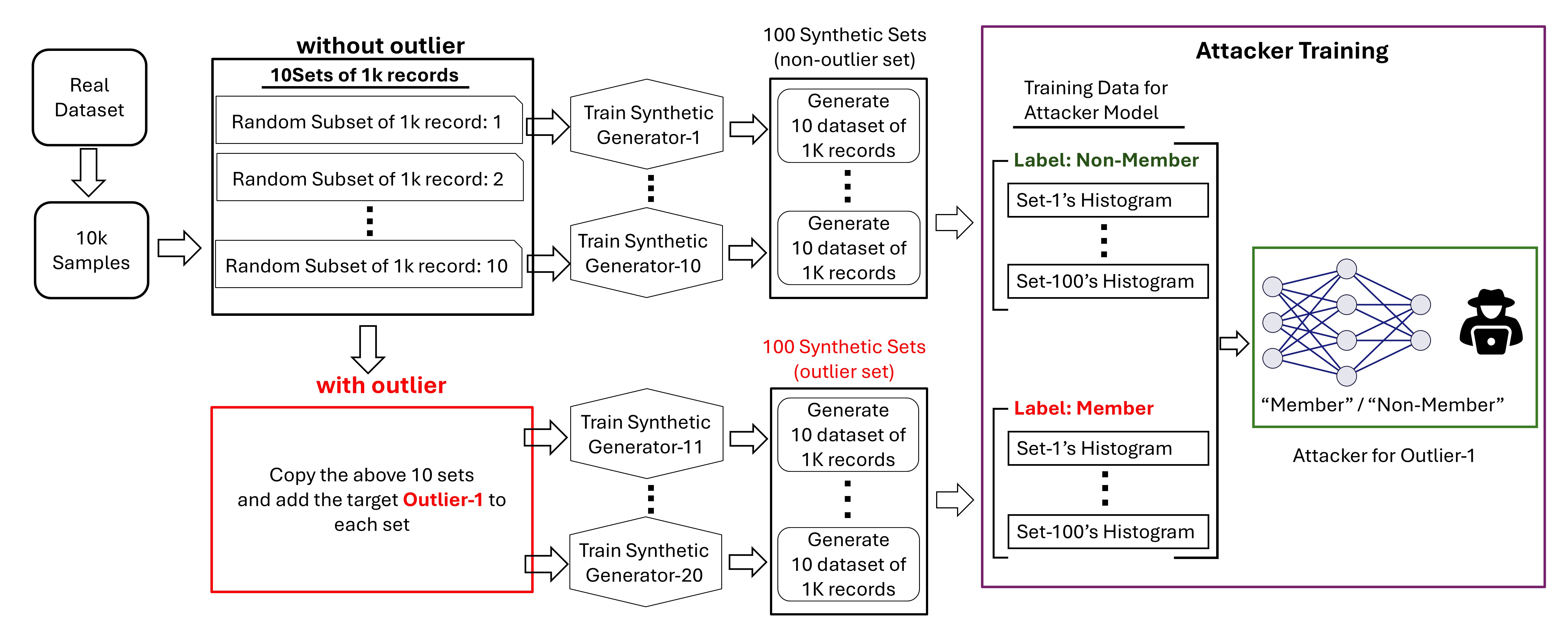} 
  \caption{An example scenario of how an attack model is trained for a specific outlier}
  \label{fig:AttackTrain}
\end{figure*}
\section*{Appendix D: Overview of SDR}\label{secGround}
In subsection D.1 and D.2, we briefly present the data publishing algorithms and the main utility metrics employed in SDR. In subsection D.3, we first discuss the novel privacy game in detail. Although this elaborate discussion may seem redundant to readers already familiar with the SDR article, it is necessary for understanding the factors that necessitated further exploration of this topic. We identify and demonstrate the specific points that limit the scope of SDR's evaluation in Section \ref{secFairComp}. 

\subsection*{D.1 Anonymization Techniques in SDR}\label{subsecSani}
\textbf{Traditional Anonymization:} SDR employed a \textit{k}-anonymization-based technique they termed ‘NHS-sanitisation’. It capped the numerical values at the 95th percentile of the respective attribute and eliminated records containing rare categories in columns. Records belonging to equivalence classes with a class size less than \textit{k} were removed. They used a fixed value of \textit{k}=10 for their \textit{k}-anonymization process. In the case of the Texas hospital inpatient discharge dataset, they selected five columns as quasi-identifiers (QIDs) out of the 18 columns. These are: the patient’s state, sex-code, race, ethnicity, and age.

\textbf{Synthetic data generation:} SDR used five synthetic data generative models in their study. Among them, two models are differentially private: PrivBayes \cite{zhang2017privbayes}, and PATE-GAN \cite{jordon2018pate}. The other three generative models are the CTGAN \cite{xu2019CTGAN}, BayNet (Bayesian networks), and IndHist \cite{ping2017indhist}. In contrast to the \textit{k-}anonymization techniques, they refrained from capping numerical values or removing rare categorical values from the raw dataset during the synthetic data generation process.

\subsection*{D.2 Utility Metrics in SDR}\label{subUTILGH}
To measure the utility of the synthetic data compared to the raw data, they employed three main approaches and used the Texas dataset for evaluation. Firstly, they used the classification accuracy to represent the machine learning utility of the published dataset (target column: \texttt{RiskMortality}). Secondly, they compared the mean and median values of the \texttt{TotalCharges}, \texttt{TotalChargesAccomm}, and \texttt{TotalChargesAncil} attributes. They also compared the marginal frequency count for the attribute \texttt{RiskMortality}. Thirdly, SDR graphically compared the impact of having a specific outlier in the training dataset on the \texttt{RiskMortality} classification of other outliers using a game. 

\subsection*{D.3 Privacy Metrics in SDR}\label{subPrivGH}
For measuring the membership-privacy of a target in the published data, SDR introduced a new metric called Privacy Gain (PG), which is calculated with the help of a novel membership inference game. 
At the heart of the privacy measurement lies a membership inference classifier trained for a specific outlier, which takes a dataset's features as input and tries to predict if that particular outlier was part of the original dataset. In other words, the membership inference of a specific outlier requires a membership classifier trained explicitly for that outlier. The metric Privacy Gain (PG) is defined as:
$$\text{PG = 1 - } \text{Adv}^{\mathrm{\textit{L}}}(\mathcal{S}, {r_t}) $$
where ${Adv}^{\mathrm{\textit{L}}}(\mathcal{S}, {r_t})$ is the attacker's advantage for a target record ${r_t}$ when a synthetic dataset {$S$} is published in place of the raw data. The attacker's advantage is implemented as $(TPR- FPR)$ where $TPR$ is the true positive rate and $FPR$ is the false positive rate of the membership classifier trained for that outlier. Below, we describe how the $TPR$ and the $FPR$ are calculated using their membership inference game.

SDR developed three distinct membership classifiers for a single target, each utilizing a different set of features. These features are:
\begin{enumerate}[noitemsep,topsep=6pt]
    \item The histogram of the attributes.
    \item The pair-wise correlation of the attributes.
    \item Simple summary statistics such as the mean, median, and variance for the numeric attributes and the count of distinct categories present along with the most frequent and least frequent category.
\end{enumerate}
If expressed in simple terms, the idea of selecting these three sets of features can be coarsely stated as follows: \textit{Based on prior experience, by examining the histograms (or the other two feature sets) of the columns in a published dataset, one can determine whether the original dataset behind the published dataset included the specific target record}. (At this point, we would like to request that the reader pay special attention to the nature of these three features. If the published datasets are truly representative of the underlying population, it seems counterintuitive that the nature of these features (e.g., frequency distribution in histogram) would allow for distinguishing between the presence or absence of a single record in the seed dataset of the published dataset. These features are not expected to vary significantly among representative samples of a population. We discuss more about this in Section \ref{secFairComp} )

For clarity and brevity, we now show the step-by-step training path of a membership attacker which utilizes the histogram feature (the same applies to the other two features as well). The evaluation (i.e., testing) process is the same, except that the membership labels are now used as the ground truth to derive the TPR and FPR of the membership classifier (i.e., the attacker). Fig. \ref{fig:AttackTrain}  depicts a simplified example of how the membership inference classifier is trained for a specific outlier. In Fig. \ref{fig:AttackTrain}, the attacker classifier is trained with 200 labeled examples (100 ‘non-member’ cases, i.e., histograms from 100 sample datasets not containing the target outlier, and histograms from 100 ‘member’ cases representing the presence of the target outlier). For the 100 ‘member’ cases, the particular target outlier whose membership the attacker wants to infer is added to the 10 datasets of 1K records,  which made them 10 samples of 1001 records each. The subsequent steps are similar to the non-member case. On the right side of the figure, within the training box, we observe the extracted feature vectors, which consist of histograms of all columns from each of the 200 datasets, used for training the attacker model for Outlier-1.

\par SDR also implemented a similar game for attribute inference attack where the attributes \texttt{Race} and \texttt{LengthOfStay} were treated as sensitive. Since the experimental results showed that BayNet and PrivBayes provide better protection than the NHS-sanitization, we did not investigate this attack further.

\section*{Appendix E: Anticipatory Questions and Answers} \label{secDisc}
In this section, we provide answers to some key questions that we anticipate may arise: 
\begin{itemize}
    \item Why might random sampling fail to represent the population accurately, despite its randomness? 
\end{itemize}
Random sampling of a small size does not always produce a representative dataset due to several reasons. 

\textbf{Sample size:} Small sample sizes are more susceptible to random variation and may not capture the full diversity of the population. The smaller the sample, the greater the chance that it will not accurately reflect the characteristics of the entire population.

\textbf{Non-uniform distribution:} If the population is not uniformly distributed, small samples are less likely to capture the variability within the population. For example, if certain characteristics are clustered in specific areas, a small sample might miss these clusters entirely.

\textbf{Lack of coverage:} Small samples might miss important subgroups within the population. For instance, rare characteristics, minority groups, or outliers may not be represented at all in a small sample, leading to incomplete or biased conclusions.

\textbf{Outliers and extreme values:} In small samples, the presence of outliers or extreme values can disproportionately influence the results, leading to skewed or biased outcomes.

\begin{itemize}
    \item Why did this study forcefully insert an outlier in the evaluation phase of the SDR (subsection \ref{subsub_test_set}) when it is already established that forcefully inserting an outlier changes the underlying data distribution? 
\end{itemize}
In subsection \ref{subsub_test_set}, our objective was not to evaluate the attacker with a representative sample. Rather, our objective was to demonstrate that if we consider the forceful insertion of an outlier as a member in the dataset as acceptable, then a similar insertion of non-member outliers causes the attacker to produce significantly more false positives, increasing the target's deniability. Since this particular attacker was trained with datasets related to the forcefully inserted outlier (a special environment), it is logical to evaluate its capabilities against a similarly forcefully inserted outlier. We acknowledge that this setting is not a very realistic environment; however, it highlights one specific limitation of the SDR's attacker.

\begin{itemize}
    \item Why do we need to consider the population for testing a membership inference attacker when the attacker is only interested in the training dataset of the generator model and a target record?  Is not the consideration for a population baseline only applicable to re-identification attacks, as shown in \cite{elemam2008k_anon}?
\end{itemize}
Note that the attacker's advantage is directly related to the difference between the true positive rate and the false positive rate. The false positive rates depends on the records from the population that were not part of the training set of the synthetic data generator. If it is not ensured that the test datasets used for evaluating an attacker's capability are representative of the population, then the attacker's capability could be overestimated or underestimated depending on the limited test data available. Therefore, it is essential to consider the distribution of population while evaluating an attacker.

\begin{itemize}
    \item Why cannot the claims based on SDR be generalized to all cases, especially considering that outliers are the most privacy-vulnerable entities in the population?
\end{itemize}

The implementation of the game in SDR resulted in an environment equivalent to a population containing only a single outlier, which is the target of the membership inference attack. Consequently, the result is not readily applicable without reservation to environments where there could be more outliers. Moreover, it is standard practice to eliminate outliers before generating synthetic datasets (even the NHS guideline that SDR followed includes this recommendation with a demonstration). Therefore, it is not appropriate to discredit the entire field of synthetic data generation based on these findings, especially since the guideline to remove outliers pre-existed for synthetic data generation.

\begin{itemize}
    \item  Why the DP bound for the privacy game is not applicable when the distributions between member and non-member dataset are significantly different?
\end{itemize}
The DP bound shown in \cite{yeom2018privacy} considers identical distributions for member and non-member datasets because if the distributions are significantly different, the attacker's advantage metric cannot distinguish how much of the privacy leakage is attributable to the model and how much is attributable to the background knowledge related to the difference between the distributions. (It can be easily shown that in some cases, the attacker's advantage is entirely attributable to distinguishing the significantly different distributions.)

\FloatBarrier
\bibliographystyle{IEEEtran}
\bibliography{mybib}   

\begin{thebibliography}{10}
\providecommand{\url}[1]{#1}
\csname url@samestyle\endcsname
\providecommand{\newblock}{\relax}
\providecommand{\bibinfo}[2]{#2}
\providecommand{\BIBentrySTDinterwordspacing}{\spaceskip=0pt\relax}
\providecommand{\BIBentryALTinterwordstretchfactor}{4}
\providecommand{\BIBentryALTinterwordspacing}{\spaceskip=\fontdimen2\font plus
\BIBentryALTinterwordstretchfactor\fontdimen3\font minus \fontdimen4\font\relax}
\providecommand{\BIBforeignlanguage}[2]{{%
\expandafter\ifx\csname l@#1\endcsname\relax
\typeout{** WARNING: IEEEtran.bst: No hyphenation pattern has been}%
\typeout{** loaded for the language `#1'. Using the pattern for}%
\typeout{** the default language instead.}%
\else
\language=\csname l@#1\endcsname
\fi
#2}}
\providecommand{\BIBdecl}{\relax}
\BIBdecl

\bibitem{fung2010privacy}
B.~C. Fung, K.~Wang, R.~Chen, and P.~S. Yu, ``Privacy-preserving data publishing: A survey of recent developments,'' \emph{ACM Computing Surveys (Csur)}, vol.~42, no.~4, pp. 1--53, 2010.

\bibitem{nist2023deid}
S.~Garfinkel, J.~Near, A.~Dajani, P.~Singer, and B.~Guttman, ``De-identifying government datasets: Techniques and governance,'' \emph{NIST Technical Series Policies}, 2023.

\bibitem{goodfellow2020generative}
I.~Goodfellow, J.~Pouget-Abadie, M.~Mirza, B.~Xu, D.~Warde-Farley, S.~Ozair, A.~Courville, and Y.~Bengio, ``Generative adversarial networks,'' \emph{Communications of the ACM}, vol.~63, no.~11, pp. 139--144, 2020.

\bibitem{kingma2013auto}
D.~P. Kingma and M.~Welling, ``Auto-encoding variational bayes,'' \emph{arXiv preprint arXiv:1312.6114}, 2013.

\bibitem{patel-NVIDIA-24}
\BIBentryALTinterwordspacing
A.~Patel, ``{NVIDIA releases open synthetic data generation pipeline for training large language models | NVIDIA blog},'' 6 2024. [Online]. Available: \url{https://blogs.nvidia.com/blog/nemotron-4-synthetic-data-generation-llm-training/}
\BIBentrySTDinterwordspacing

\bibitem{jordon2022synthetic}
J.~Jordon, L.~Szpruch, F.~Houssiau, M.~Bottarelli, G.~Cherubin, C.~Maple, S.~N. Cohen, and A.~Weller, ``Synthetic data--what, why and how?'' \emph{arXiv preprint arXiv:2205.03257}, 2022.

\bibitem{cai2022GANs}
\BIBentryALTinterwordspacing
Z.~Cai, Z.~Xiong, H.~Xu, P.~Wang, W.~Li, and Y.~Pan, ``Generative adversarial networks: A survey toward private and secure applications,'' \emph{ACM Computing Surveys (CSUR)}, vol.~54, no.~6, 2021. [Online]. Available: \url{https://doi.org/10.1145/3459992}
\BIBentrySTDinterwordspacing

\bibitem{hernandez2022synthetic}
M.~Hernandez, G.~Epelde, A.~Alberdi, R.~Cilla, and D.~Rankin, ``Synthetic data generation for tabular health records: A systematic review,'' \emph{Neurocomputing}, vol. 493, pp. 28--45, 2022.

\bibitem{elemam2020practical}
K.~El~Emam, L.~Mosquera, and R.~Hoptroff, \emph{Practical synthetic data generation: balancing privacy and the broad availability of data}.\hskip 1em plus 0.5em minus 0.4em\relax USA: O'Reilly Media Inc., 2020.

\bibitem{mcclure2012differential}
D.~McClure and J.~P. Reiter, ``Differential privacy and statistical disclosure risk measures: An investigation with binary synthetic data.'' \emph{Trans. Data Priv.}, vol.~5, no.~3, pp. 535--552, 2012.

\bibitem{liu2022synth}
\BIBentryALTinterwordspacing
F.~Liu, Z.~Cheng, H.~Chen, Y.~Wei, L.~Nie, and M.~Kankanhalli, ``Privacy-preserving synthetic data generation for recommendation systems,'' in \emph{Proceedings of the 45th International ACM SIGIR Conference on Research and Development in Information Retrieval}, ser. SIGIR '22.\hskip 1em plus 0.5em minus 0.4em\relax New York, NY, USA: Association for Computing Machinery, 2022, p. 1379–1389. [Online]. Available: \url{https://doi.org/10.1145/3477495.3532044}
\BIBentrySTDinterwordspacing

\bibitem{elemam2020eval}
K.~El~Emam, L.~Mosquera, and J.~Bass, ``Evaluating identity disclosure risk in fully synthetic health data: model development and validation,'' \emph{Journal of medical Internet research}, vol.~22, no.~11, p. e23139, 2020.

\bibitem{zhang2021privsyn}
Z.~Zhang, T.~Wang, N.~Li, J.~Honorio, M.~Backes, S.~He, J.~Chen, and Y.~Zhang, ``$\{$PrivSyn$\}$: Differentially private data synthesis,'' in \emph{30th USENIX Security Symposium (USENIX Security 21)}.\hskip 1em plus 0.5em minus 0.4em\relax USENIX Association, 2021, pp. 929--946.

\bibitem{samarati1998protecting}
P.~Samarati and L.~Sweeney, ``Protecting privacy when disclosing information: k-anonymity and its enforcement through generalization and suppression,'' \emph{Technical Report, SRI-CSL}, vol. SRI-CSL-98-04, pp. 1--19, 1998.

\bibitem{aggarwal2005k}
C.~C. Aggarwal, ``On k-anonymity and the curse of dimensionality,'' in \emph{VLDB}, vol.~5, 2005, pp. 901--909.

\bibitem{elemam2008k_anon}
\BIBentryALTinterwordspacing
K.~El~Emam and F.~K. Dankar, ``{Protecting Privacy Using k-Anonymity},'' \emph{Journal of the American Medical Informatics Association}, vol.~15, no.~5, pp. 627--637, 09 2008. [Online]. Available: \url{https://doi.org/10.1197/jamia.M2716}
\BIBentrySTDinterwordspacing

\bibitem{Angiuli2015kSkew}
\BIBentryALTinterwordspacing
O.~Angiuli, J.~Blitzstein, and J.~Waldo, ``How to de-identify your data: Balancing statistical accuracy and subject privacy in large social-science data sets,'' \emph{ACM Queue}, vol.~13, no.~8, p. 20–39, sep 2015. [Online]. Available: \url{https://doi.org/10.1145/2838344.2838930}
\BIBentrySTDinterwordspacing

\bibitem{slijepvcevic2021k}
D.~Slijepvcevic, M.~Henzl, L.~D. Klausner, T.~Dam, P.~Kieseberg, and M.~Zeppelzauer, ``k-anonymity in practice: How generalisation and suppression affect machine learning classifiers,'' \emph{Computers and Security}, vol. 111, p. 102488, 2021.

\bibitem{machanavajjhala2007diversity}
A.~Machanavajjhala, D.~Kifer, J.~Gehrke, and M.~Venkitasubramaniam, ``l-diversity: Privacy beyond k-anonymity,'' \emph{ACM Transactions on Knowledge Discovery from Data (TKDD)}, vol.~1, no.~1, pp. 3--es, 2007.

\bibitem{stadler2022synthetic}
T.~Stadler, B.~Oprisanu, and C.~Troncoso, ``Synthetic data--anonymisation groundhog day,'' in \emph{31st USENIX Security Symposium (USENIX Security 22)}.\hskip 1em plus 0.5em minus 0.4em\relax Boston, MA: USENIX Association, 2022, pp. 1451--1468.

\bibitem{ground2021git}
E.~SPRING~Laboratory, ``synthetic data release,'' \emph{https://github.com/spring-epfl/synthetic\_data\_release}, pp. Accessed 2023--10--17, 2021.

\bibitem{jordon2018pate}
\BIBentryALTinterwordspacing
J.~Jordon, J.~Yoon, and M.~Van Der~Schaar, ``Pate-gan: Generating synthetic data with differential privacy guarantees,'' in \emph{International conference on learning representations}.\hskip 1em plus 0.5em minus 0.4em\relax New Orleans, Louisiana, United States: OpenReview.net, 2019, pp. 1--21. [Online]. Available: \url{https://openreview.net/forum?id=S1zk9iRqF7}
\BIBentrySTDinterwordspacing

\bibitem{zhang2017privbayes}
J.~Zhang, G.~Cormode, C.~M. Procopiuc, D.~Srivastava, and X.~Xiao, ``Privbayes: Private data release via bayesian networks,'' \emph{ACM Transactions on Database Systems (TODS)}, vol.~42, no.~4, pp. 1--41, 2017.

\bibitem{yeom2018privacy}
\BIBentryALTinterwordspacing
S.~Yeom, I.~Giacomelli, M.~Fredrikson, and S.~Jha, ``Privacy risk in machine learning: Analyzing the connection to overfitting,'' \emph{arXiv:1709.01604v5 [cs.CR]}, 2018. [Online]. Available: \url{https://arxiv.org/pdf/1709.01604.pdf}
\BIBentrySTDinterwordspacing

\bibitem{sweeney2002k}
L.~Sweeney, ``k-anonymity: A model for protecting privacy,'' \emph{International journal of uncertainty, fuzziness and knowledge-based systems}, vol.~10, no.~05, pp. 557--570, 2002.

\bibitem{li2006t}
N.~Li, T.~Li, and S.~Venkatasubramanian, ``t-closeness: Privacy beyond k-anonymity and l-diversity,'' in \emph{2007 IEEE 23rd international conference on data engineering}, IEEE.\hskip 1em plus 0.5em minus 0.4em\relax Istanbul, Turkey: IEEE, 2006, pp. 106--115.

\bibitem{lefevre2006mondrian}
K.~LeFevre, D.~J. DeWitt, and R.~Ramakrishnan, ``Mondrian multidimensional k-anonymity,'' in \emph{22nd International conference on data engineering (ICDE'06)}, IEEE.\hskip 1em plus 0.5em minus 0.4em\relax Atlanta, GA, USA: IEEE, 2006, pp. 25--25.

\bibitem{lefevre2005incognito}
------, ``Incognito: Efficient full-domain k-anonymity,'' in \emph{Proceedings of the 2005 ACM SIGMOD international conference on Management of data}.\hskip 1em plus 0.5em minus 0.4em\relax New York, NY, USA: Association for Computing Machinery, 2005, pp. 49--60.

\bibitem{sweeney1998datafly}
L.~Sweeney, ``Datafly: A system for providing anonymity in medical data,'' \emph{Database Security XI: Status and Prospects}, pp. 356--381, 1998.

\bibitem{samarati2001protecting}
P.~Samarati, ``Protecting respondents identities in microdata release,'' \emph{IEEE transactions on Knowledge and Data Engineering}, vol.~13, no.~6, pp. 1010--1027, 2001.

\bibitem{bayardo2005data}
R.~J. Bayardo and R.~Agrawal, ``Data privacy through optimal k-anonymization,'' in \emph{21st International conference on data engineering (ICDE'05)}, IEEE.\hskip 1em plus 0.5em minus 0.4em\relax Tokyo, Japan: IEEE, 2005, pp. 217--228.

\bibitem{wang2007handicapping}
K.~Wang, B.~C. Fung, and P.~S. Yu, ``Handicapping attacker's confidence: an alternative to k-anonymization,'' \emph{Knowledge and Information Systems}, vol.~11, pp. 345--368, 2007.

\bibitem{meyerson2004complexity}
A.~Meyerson and R.~Williams, ``On the complexity of optimal k-anonymity,'' in \emph{Proceedings of the twenty-third ACM SIGMOD-SIGACT-SIGART symposium on Principles of database systems}, 2004, pp. 223--228.

\bibitem{rubin1993statistical}
D.~B. Rubin, ``Statistical disclosure limitation,'' \emph{Journal of official Statistics}, vol.~9, no.~2, pp. 461--468, 1993.

\bibitem{young2009using}
J.~Young, P.~Graham, and R.~Penny, ``Using bayesian networks to create synthetic data,'' \emph{Journal of Official Statistics}, vol.~25, no.~4, pp. 549--567, 2009.

\bibitem{ngoko2014synthetic}
B.~Ngoko, H.~Sugihara, and T.~Funaki, ``Synthetic generation of high temporal resolution solar radiation data using markov models,'' \emph{Solar Energy}, vol. 103, pp. 160--170, 2014.

\bibitem{xie2018differentially}
L.~Xie, K.~Lin, S.~Wang, F.~Wang, and J.~Zhou, ``Differentially private generative adversarial network,'' \emph{arXiv preprint arXiv:1802.06739}, 2018.

\bibitem{wagner2018pvmetrics}
I.~Wagner and D.~Eckhoff, ``Technical privacy metrics: a systematic survey,'' \emph{ACM Computing Surveys (CSUR)}, vol.~51, no.~3, pp. 1--38, 2018.

\bibitem{dwork2017attacks}
\BIBentryALTinterwordspacing
C.~Dwork, A.~Smith, T.~Steinke, and J.~Ullman, ``Exposed! a survey of attacks on private data,'' \emph{Annual Review of Statistics and Its Application}, vol.~4, no.~1, pp. 61--84, 2017. [Online]. Available: \url{https://doi.org/10.1146/annurev-statistics-060116-054123}
\BIBentrySTDinterwordspacing

\bibitem{giomi2022unified}
M.~Giomi, F.~Boenisch, C.~Wehmeyer, and B.~Tasn{\'a}di, ``A unified framework for quantifying privacy risk in synthetic data,'' \emph{arXiv preprint arXiv:2211.10459}, 2022.

\bibitem{el2011systematic}
K.~El~Emam, E.~Jonker, L.~Arbuckle, and B.~Malin, ``A systematic review of re-identification attacks on health data,'' \emph{PloS one}, vol.~6, no.~12, p. e28071, 2011.

\bibitem{merener2012linkage}
M.~M. Merener, ``Theoretical results on de-anonymization via linkage attacks.'' \emph{Trans. Data Priv.}, vol.~5, no.~2, pp. 377--402, 2012.

\bibitem{shokri2017membership}
R.~Shokri, M.~Stronati, C.~Song, and V.~Shmatikov, ``Membership inference attacks against machine learning models,'' in \emph{2017 IEEE symposium on security and privacy (SP)}.\hskip 1em plus 0.5em minus 0.4em\relax IEEE, 2017, pp. 3--18.

\bibitem{salem2018ml}
A.~Salem, Y.~Zhang, M.~Humbert, P.~Berrang, M.~Fritz, and M.~Backes, ``Ml-leaks: Model and data independent membership inference attacks and defenses on machine learning models,'' \emph{arXiv preprint}, vol. arXiv:1806.01246, 2018.

\bibitem{carlini2022membership}
N.~Carlini, S.~Chien, M.~Nasr, S.~Song, A.~Terzis, and F.~Tramer, ``Membership inference attacks from first principles,'' in \emph{2022 IEEE Symposium on Security and Privacy (SP)}, IEEE.\hskip 1em plus 0.5em minus 0.4em\relax San Francisco, CA, USA, 2022, pp. 1897--1914.

\bibitem{tf2020tfmia}
``Tensorflow privacy: Library for training machine learning models with privacy for training data,'' 2023, available at: \url{https://github.com/tensorflow/privacy} Accessed: 24th July 2023.

\bibitem{ganev2023inadequacy}
G.~Ganev and E.~De~Cristofaro, ``On the inadequacy of similarity-based privacy metrics: Reconstruction attacks against "truly anonymous synthetic data",'' \emph{arXiv preprint arXiv:2312.05114}, 2023.

\bibitem{powar23petsattrib}
J.~Powar and A.~R. Beresford, ``Sok: Managing risks of linkage attacks on data privacy,'' \emph{Proceedings on Privacy Enhancing Technologies}, vol.~2, pp. 97--116, 2023.

\bibitem{jia2018attriguard}
J.~Jia and N.~Z. Gong, ``$\{$AttriGuard$\}$: A practical defense against attribute inference attacks via adversarial machine learning,'' in \emph{27th USENIX Security Symposium (USENIX Security 18)}, 2018, pp. 513--529.

\bibitem{jayaraman2022attribute}
B.~Jayaraman and D.~Evans, ``Are attribute inference attacks just imputation?'' in \emph{Proceedings of the 2022 ACM SIGSAC Conference on Computer and Communications Security}, 2022, pp. 1569--1582.

\bibitem{fredrikson2014warfarin}
M.~Fredrikson, E.~Lantz, S.~Jha, S.~Lin, D.~Page, and T.~Ristenpart, ``Privacy in pharmacogenetics: An $\{$End-to-End$\}$ case study of personalized warfarin dosing,'' in \emph{23rd USENIX security symposium (USENIX Security 14)}, 2014, pp. 17--32.

\bibitem{lefevre2008workload}
K.~LeFevre, D.~J. DeWitt, and R.~Ramakrishnan, ``Workload-aware anonymization techniques for large-scale datasets,'' \emph{ACM Transactions on Database Systems (TODS)}, vol.~33, no.~3, pp. 1--47, 2008.

\bibitem{fung2007anonymizing}
B.~C. Fung, K.~Wang, and S.~Y. Philip, ``Anonymizing classification data for privacy preservation,'' \emph{IEEE transactions on knowledge and data engineering}, vol.~19, no.~5, pp. 711--725, 2007.

\bibitem{NHS2019synth}
\BIBentryALTinterwordspacing
A.~Services, ``Attendances and emergency (a\&e) synthetic data,'' \emph{NHS Data Catalogue}, pp. Accessed 2023--10--17, Oct 2019. [Online]. Available: \url{https://data.england.nhs.uk/dataset/a-e-synthetic-data}
\BIBentrySTDinterwordspacing

\bibitem{LuiP2015outLpriv}
\BIBentryALTinterwordspacing
E.~Lui and R.~Pass, ``Outlier privacy,'' in \emph{Theory of Cryptography - 12th Theory of Cryptography Conference, {TCC} March 23-25, 2015, Proceedings, Part {II}}, ser. Lecture Notes in Computer Science, vol. 9015.\hskip 1em plus 0.5em minus 0.4em\relax Warsaw, Poland: Springer, 2015, pp. 277--305. [Online]. Available: \url{https://doi.org/10.1007/978-3-662-46497-7\_11}
\BIBentrySTDinterwordspacing

\bibitem{mayer2020privacy}
R.~Mayer, M.~Hittmeir, and A.~Ekelhart, ``Privacy-preserving anomaly detection using synthetic data,'' in \emph{Data and Applications Security and Privacy XXXIV: 34th Annual IFIP WG 11.3 Conference, DBSec 2020, Proceedings 34}, Springer.\hskip 1em plus 0.5em minus 0.4em\relax Regensburg, Germany: Springer, 2020, pp. 195--207.

\bibitem{hu2024sok}
Y.~Hu, F.~Wu, Q.~Li, Y.~Long, G.~M. Garrido, C.~Ge, B.~Ding, D.~Forsyth, B.~Li, and D.~Song, ``Sok: Privacy-preserving data synthesis,'' in \emph{2024 IEEE Symposium on Security and Privacy (SP)}.\hskip 1em plus 0.5em minus 0.4em\relax IEEE, 2024, pp. 4696--4713.

\bibitem{ping2017datasynthesizer}
H.~Ping, J.~Stoyanovich, and B.~Howe, ``Datasynthesizer: Privacy-preserving synthetic datasets,'' in \emph{Proceedings of the 29th International Conference on Scientific and Statistical Database Management}, 2017, pp. 1--5.

\bibitem{kotelnikov2023tabddpm}
A.~Kotelnikov, D.~Baranchuk, I.~Rubachev, and A.~Babenko, ``Tabddpm: Modelling tabular data with diffusion models,'' in \emph{International Conference on Machine Learning}.\hskip 1em plus 0.5em minus 0.4em\relax PMLR, 2023, pp. 17\,564--17\,579.

\bibitem{baowaly2019synthesizing}
M.~K. Baowaly, C.-C. Lin, C.-L. Liu, and K.-T. Chen, ``Synthesizing electronic health records using improved generative adversarial networks,'' \emph{Journal of the American Medical Informatics Association}, vol.~26, no.~3, pp. 228--241, 2019.

\bibitem{tao2021benchmarking}
Y.~Tao, R.~McKenna, M.~Hay, A.~Machanavajjhala, and G.~Miklau, ``Benchmarking differentially private synthetic data generation algorithms,'' \emph{AAAI Workshop on Privacy-Preserving Artificial Intelligence (PPAI-22)}, 2022.

\bibitem{SDV}
N.~Patki, R.~Wedge, and K.~Veeramachaneni, ``The synthetic data vault,'' in \emph{IEEE International Conference on Data Science and Advanced Analytics (DSAA)}.\hskip 1em plus 0.5em minus 0.4em\relax Montreal, QC, Canada: IEEE, Oct 2016, pp. 399--410.

\bibitem{Schaar2023Synth}
\BIBentryALTinterwordspacing
Z.~Qian, B.-C. Cebere, and M.~van~der Schaar, ``Synthcity: facilitating innovative use cases of synthetic data in different data modalities,'' 2023. [Online]. Available: \url{https://arxiv.org/abs/2301.07573}
\BIBentrySTDinterwordspacing

\bibitem{hodges1958kstSig}
J.~Hodges~Jr, ``The significance probability of the smirnov two-sample test,'' \emph{Arkiv f{\"o}r matematik}, vol.~3, no.~5, pp. 469--486, 1958.

\bibitem{TVcomplement}
\BIBentryALTinterwordspacing
``Tvcomplement - sdmetrics,'' 2023, available at: \url{https://docs.sdv.dev/sdmetrics/metrics/metrics-glossary/tvcomplement} Accessed: 24th July 2023. [Online]. Available: \url{https://docs.sdv.dev/sdmetrics/metrics/metrics-glossary/tvcomplement}
\BIBentrySTDinterwordspacing

\bibitem{tang2022mitigating}
X.~Tang, S.~Mahloujifar, L.~Song, V.~Shejwalkar, M.~Nasr, A.~Houmansadr, and P.~Mittal, ``Mitigating membership inference attacks by $\{$Self-Distillation$\}$ through a novel ensemble architecture,'' in \emph{31st USENIX Security Symposium (USENIX Security 22)}, 2022, pp. 1433--1450.

\bibitem{song2021systematic}
L.~Song and P.~Mittal, ``Systematic evaluation of privacy risks of machine learning models,'' in \emph{30th USENIX Security Symposium (USENIX Security 21)}, 2021, pp. 2615--2632.

\bibitem{maritsch2022data}
F.~Maritsch, I.~Cil, C.~McKinnon, J.~Potash, N.~Baumgartner, V.~Philippon, and B.~G. Pavlova, ``Data privacy protection in scientific publications: process implementation at a pharmaceutical company,'' \emph{BMC medical ethics}, vol.~23, no.~1, p.~65, 2022.

\bibitem{branson2020evaluating}
J.~Branson, N.~Good, J.-W. Chen, W.~Monge, C.~Probst, and K.~El~Emam, ``Evaluating the re-identification risk of a clinical study report anonymized under ema policy 0070 and health canada regulations,'' \emph{Trials}, vol.~21, pp. 1--9, 2020.

\bibitem{adult2}
B.~Becker and R.~Kohavi, ``{Adult},'' UCI Machine Learning Repository, 1996, {DOI}: https://doi.org/10.24432/C5XW20.

\bibitem{CreditCardChurn}
\BIBentryALTinterwordspacing
``Credit card customers,'' accessed: 29th September 2024. [Online]. Available: \url{https://www.kaggle.com/datasets/sakshigoyal7/credit-card-customers}
\BIBentrySTDinterwordspacing

\bibitem{xu2019CTGAN}
L.~Xu, M.~Skoularidou, A.~Cuesta-Infante, and K.~Veeramachaneni, ``Modeling tabular data using conditional gan,'' \emph{Proceedings of the 33rd International Conference on Neural Information Processing Systems}, vol. Article 659, pp. 7335–--7345, 2019.

\bibitem{ping2017indhist}
H.~Ping, J.~Stoyanovich, and B.~Howe, ``Datasynthesizer: Privacy-preserving synthetic datasets,'' in \emph{Proceedings of the 29th International Conference on Scientific and Statistical Database Management}.\hskip 1em plus 0.5em minus 0.4em\relax New York, NY, USA: Association for Computing Machinery, 2017, pp. 1--5.

\end{thebibliography}
\end{document}